\documentclass{pasj00}
\SetRunningHead{Star Formation Law in IC 342}{Pan, Kuno, and Hirota}

\usepackage{times}
\usepackage{graphicx}
\usepackage{threeparttable}

\begin{document}

\title{Environmental Dependence of Star Formation Law in the Disk and Center of IC 342}
\author{Hsi-An Pan\altaffilmark{1,2}, Nario Kuno\altaffilmark{1,2}, and Akihiko Hirota\altaffilmark{2}}
\affil{$^{1}$Department of Astronomical Science, The Graduate University for Advanced Studies, Shonan Village, Hayama, Kanagawa 240-0193, Japan}
\affil{$^{2}$Nobeyama Radio Observatory, NAOJ, Minamimaki, Minamisaku, Nagano 384-1305, Japan}

\email{pan.h.a@nao.ac.jp}
\KeyWords{galaxies: individual (IC 342)${}$ --- galaxies: ISM${}$ --- ISM: molecules${}$}

\maketitle

\begin{abstract}

The Kennicutt-Schmidt (K--S) law in IC 342 is examined using the $^{12}$CO-to-H$_{2}$ conversion factor ($X_{\mathrm{CO,v}}$), which depends on the metallicity and CO intensity. Additionally, an optically thin $^{13}$CO (1-0) is also independently used to analyze the K--S law. 
$X_{\mathrm{CO,v}}$ is two to three times lower than the Galactic standard $X_{\mathrm{CO}}$ in the galactic center and approximately two  times higher than $X_{\mathrm{CO}}$ at the disk.
The surface densities of molecular gas ($\Sigma _{\mathrm{H_{2}}}$) derived from $^{12}$CO and $^{13}$CO are consistent at the environment in a high-$\Sigma _{\mathrm{H_{2}}}$ region. 
By comparing the K-S law in the disk and the central regions of IC 342, we found that the power law index of K-S law ($N$) increases toward the central region.
Furthermore, the dependence of $N$ on $\Sigma _{\mathrm{H_{2}}}$ is observed.
Specifically, $N$ increases with  $\Sigma _{\mathrm{H_{2}}}$.
The derived $N$ in this work and  previous observations are consistent with the implication that star formation is likely triggered by gravitational instability in the disk (low-$\Sigma _{\mathrm{H_{2}}}$ region) of IC 342 and both gravitational instability and cloud-cloud collisions in the central region (high-$\Sigma _{\mathrm{H_{2}}}$ regime).
In addition, the increasing $N$ toward the high-$\Sigma _{\mathrm{H_{2}}}$ domain also matches the theoretical prediction regarding the properties of giant molecular clouds.  
The results of IC 342 are supported by the same analysis of other nearby galaxies.

\end{abstract}

\section{Introduction}

\citet{Sch59} and \citet{Ken98} have proposed an empirical star formation  law between the surface densities of the star formation rate ($\Sigma _{\mathrm{SFR}}$) and the total gas ($\Sigma _{\mathrm{gas}}$),  widely known as the Kennicutt-Schmidt (K--S) law. The K--S law is written in the form of  
\begin{equation}
\Sigma _{\mathrm{SFR}} \mathrm{(M_{\solar} yr^{-1} kpc^{-2})} \propto \Sigma _{\mathrm{gas}}^{N}  \mathrm{(M_{\solar} pc^{-2})}.
\end{equation}
As stars are formed from molecular gas instead of atomic gas, $\Sigma _{\mathrm{gas}}$ in the K--S law is often replaced with $\Sigma _{\mathrm{H_{2}}}$.

With $^{12}$CO observations and the Galactic standard CO-to-H$_{2}$ conversion factor ($X_{\mathrm{CO}}$ $=$ 2.3  $\times$ 10$^{20}$ cm$^{-2}$ (K km s$^{-1}$)$^{-1}$; \cite{Str88}), numerous works show that $N$ is in the range of 1.0 -- 2.0. 
The slope of the K--S law may provide us with clues that will help to explain the mechanism of star formation.
For example, if star formation only depends on the mass of giant molecular clouds (GMCs),  a constant SFR per unit mass of molecular gas is expected, leading to a linear relation between $\Sigma _{\mathrm{SFR}} $ and $\Sigma _{\mathrm{H_{2}}}$ and, namely, $N\approx1.0$ \citep{Big08,Dob09}. 
If molecular gas is converted to stars with a timescale of free-fall time, and the scale height of gas is approximately constant everywhere in a galaxy, a slope of 1.5 is expected \citep{Elm94,Elm02,Li06}.
If cloud-cloud collisions trigger star formation, the star formation timescale is set by the collision rate of the virially bound objects, resulting in a slope of $\sim$2.0 \citep{Tan00,Tas09}.

There are, however, some drawbacks of the current method of deriving the K--S law.
At first, $X_{\mathrm{CO}}$ is not a constant but a function of metallicity and CO intensity \citep{Isr97,Gen12,Nar12}.
Observations have also shown that $X_{\mathrm{CO}}$ increases with decreasing oxygen abundance \citep{Dic86,Wil95, Ari96} and CO intensity \citep{Nak95}. Because oxygen abundance appears to be a function of the galactic radius in spiral galaxies \citep{Pil04,Mou10}, $X_{\mathrm{CO}}$ may vary within a single galaxy as well. For example, $X_{\mathrm{CO}}$ is 0.75  $\times$ 10$^{20}$  cm$^{-2}$ (K km s$^{-1}$)$^{-1}$ within the inner 1$\farcs$2 of M51. It is approximately three times lower than the Galactic standard $X_{\mathrm{CO}}$, while its value rises to (1 -- 4) $\times$ 10$^{20}$ cm$^{-2}$ (K km s$^{-1}$)$^{-1}$ at $>$ 2$\farcs$0 \citep{Nak95}.
For this reason, applying the Galactic standard $X_{\mathrm{CO}}$ to external galaxies is not ideal. 
Secondly, the nature of $^{12}$CO also limits its ability to determine $\Sigma _{\mathrm{H_{2}}}$. 
$^{12}$CO is optically thick, so it traces the emissions from the envelope of the GMCs, rather than the dense clumps/cores in the GMCs, where stars are formed. Likewise, because of its high opacity, the intensity of $^{12}$CO does not indicate the amount of gas when GMCs start to pile up along the line of sight. Furthermore, because $^{12}$CO can be excited at a density as low as $\sim$ 100 cm$^{-3}$ 
as a result of the photon trapping effect, the gas traced by $^{12}$CO is not necessarily associated with star formation. 
Finally,  diffuse emissions that do not relate to star formation are involved in the tracers of SFR, such as UV, 24$\mu$m, and H$\alpha$.
Although \citet{Liu11} show that the presence of diffuse emissions can alter the relation of $\Sigma _{\mathrm{H_{2}}}$ and $\Sigma _{\mathrm{SFR}}$  from linear to nonlinear, \citet{Rah11} concluded that the contribution from the diffuse emissions alone is not sufficient to explain all the differences in the observed slopes.  

Inspired by the above issues,  we study the K--S law in a nearby galaxy IC 342 in terms of two aspects: the CO-to-H$_{2}$ conversion factors which depend on the metallicity and CO intensity \citep{Nak95,Nar12}, and the utilization of an optically thin line $^{13}$CO (1--0), which is reckoned to be a better tracer of the real mass distribution (e.g., \cite{Aal10}).
IC 342 is a nearby face-on barred spiral galaxy located at 3.3 Mpc. 1$\arcsec$ corresponds to 16 pc at that distance \citep{Sah02}. 
The CO map made with a single dish  shows a clear molecular bar with a length of approximately 4.5 kpc \citep{Cro01}. 
A primary spiral arm emerging from the southern bar is found in the CO map. 
IC 342 is undergoing bar-driven gas inflow \citep{Sch03} which triggers the starburst at the center and therefore splits the physical environment from the galactic disk, providing a chance to study the K--S law in diverse physical conditions of molecular gas.

The outline of this paper is as follows: the radio and infrared published data we used in this work are presented in Section \ref{sec_data}. The study of the radial distribution of the gas and SFR is shown in Section \ref{RadDist}. The methods of deriving $\Sigma _{\mathrm{H_{2}}}$ from $^{12}$CO (1--0) and $^{13}$CO (1--0) are demonstrated in Section \ref{Sec_Xcolow} and Section \ref{KS13_high}, respectively. The uncertainties of the methods are discussed in Section \ref{UncerMass}. Section \ref{KS} presents the derived K--S law with both $^{12}$CO (1--0) and $^{13}$CO (1--0).
In Section \ref{galaxies_literatures}, we examine the K-S law in other nearby galaxies using the published data of $^{12}$CO and metallicity to confirm the results from IC 342.    
The probable mechanisms of star formation in IC 342 are considered in Section \ref{SF_IC342}.

\section{Data}
\label{sec_data}
\subsection{NRO CO Maps}
\begin{figure}
\begin{center}
\FigureFile(80mm,80mm){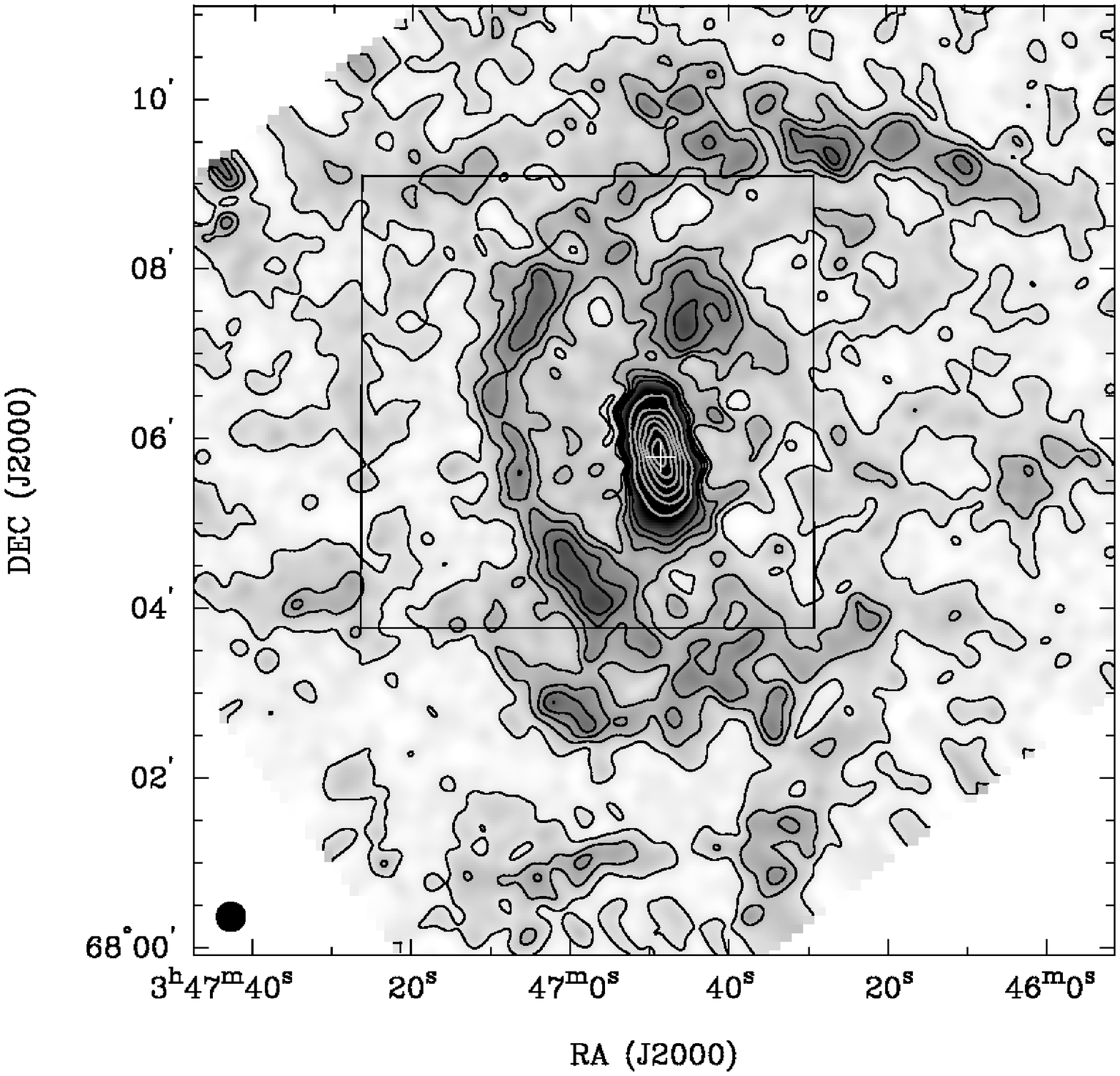}
\end{center}
\caption{$^{12}$CO map of IC 342 obtained from \citet{Kun07}. Contours represent 4.5 (3 $\sigma$) 10, 15, 20, 25, 50, 80, 110, 140, 170, and 200 K km s$^{-1}$. The white cross marks the galactic center. The beam size of 20$\arcsec$ is shown in the bottom left of the figure. The area of the $^{13}$CO (1-0) observation made by \citet{Hir10} is indicated with a box.}
\label{FIG_ico}
\end{figure}

The $^{12}$CO ($J\,= \, 1\rightarrow 0$) (hereafter $^{12}$CO) image of IC 342 (Figure \ref{FIG_ico}) was taken from the Nobeyama CO Atlas of Nearby Spiral Galaxies \citep{Kun07}.  
The observations were made by the Nobeyama 45-m telescope. The angular resolution of the map is 20$\arcsec$ at 115 GHz, corresponding to a physical resolution of $\sim$320 pc. 
The map extends up to a galactocentric radius of approximately 6.5 kpc (406$\arcsec$).
The rms noise is approximately 1.5 K km s$^{-1}$ or $\sim$ 5 M$_{\solar}$ pc$^{-2}$ with the Galactic standard CO-to-H$_{2}$ conversion factor ($X_{\mathrm{CO}}$ $=$  2.3 $\times$ 10$^{20}$ cm$^{-2}$ (K km s$^{-1}$)$^{-1}$) \citep{Str88}.  

The $^{13}$CO  ($J\,= \, 1\rightarrow 0$) (hereafter $^{13}$CO) image (yellow contours in Figure \ref{FIG_structure}) was also taken with the Nobeyama 45-m telescope by \citet{Hir10}.
The map covers the galactic center, the bar, and the spiral arm emerging from the southern bar end with a total area of $\sim$320$\arcsec$ $\times$320 $\arcsec$ (5.1 kpc $\times$ 5.1 kpc), which is approximately one fourth of the $^{12}$CO map.
The angular resolution is also 20$\arcsec$ (320 pc) at 110 GHz and the rms noise is 0.4 K km s$^{-1}$.

\subsection{VLA HI Map}
\label{sec_data_HI}
The atomic hydrogen (HI) image of IC 342 is taken from \citet{Cro00}. The observations were carried out by the VLA (Very Large Array) using the C and D configurations. The resulting angular resolution was 38$\arcsec$ $\times$ 37$\arcsec$ (610 pc $\times$ 590 pc) at the wavelength of 21 cm. 
The largest structure the observation sensitive to is about 15$\arcmin$. The amount of missing flux due to the short spacing problem was approximately 30\%  \citep{Cro00}.
The HI image has a radius of 1300$\arcsec$ ($\sim$21 kpc), which is far beyond the molecular disk of IC 342 by approximately a factor of three.

\subsection{Spitzer 24 $\mu$m Map}
\label{sec_data_sp}
IC 342 is not included in the Spitzer Local Volume Legacy Survey (LVL) and the Spitzer Infrared Nearby Galaxies Survey (SINGS), but in several individual programs, presumably because of its low latitude ($b$=10$\fdg$58).
We used the 24$\mu$m data  from Spitzer Heritage Archive (SHA) as SFR tracer. The data were obtained by the Multiband Imaging Photometer (MIPS) \citep{Rie04}. The angular resolution of  the 24 $\mu$m map is $\sim$5$\farcs$7 ($\sim$ 91 pc).

\subsection{Hershel Mid- and Far-Infrared Maps}
\label{sec_data_her}
The Hershel telescope provides infrared data from 70 to 500$\mu$m. The data are useful to derive the dust/gas temperature from the spectral energy distribution (SED), which is required in our analysis  (\S\ref{KS13_high}). 
IC 342 is one of the samples in the project of Key Insights on Nearby Galaxies: a Far-Infrared Survey with Herschel (KINGFISH) \citep{Ken11}. The data are available on the project website.
Herschel PACS and SPIRE instruments provide the data at  wavelengths of  70, 100, and 160 $\mu$m (PACS) and 250, 350, and 500 $\mu$m (SPIRE). 
The original units of the images, the pixel sizes and the resolutions are listed in Table \ref{IR_instrument}.  The process of data reduction is described in the KINGFISH Data Products Delivery Users Guide.

\section{Radial Distribution of Gas and Star Formation}
\label{RadDist}
\subsection{Radial Distribution of Gas}
The neutral gas in a galaxy consists of molecular and atomic gas. The former is required for star formation, but some nearby spiral galaxies harbor larger amounts of atomic gas relative to the molecular gas.  Accordingly, we checked the content of the gas in IC 342 as the first step.

We use the image of $^{12}$CO to measure the radial distribution of $\Sigma _{\mathrm{H_{2}}}$. 
The inclination and the position angle of the galaxy are 31$^{\circ}$ and 37$^{\circ}$, respectively \citep{Cro00}.
The step size between each azimuthal average data point is 10$\arcsec$, starting from 0$\arcsec$ to 400$\arcsec$ (6.4 kpc) or R/R$_{25}$ $\approx$ 0.30, where R$_{25}$ $=$ 22.3$\arcmin$ or 21.4 kpc \citep{Pil04}.
$X_{\mathrm{CO}}$ of  2.3 $\times$ 10$^{20}$ cm$^{-2}$ (K km s$^{-1}$)$^{-1}$ \citep{Str88} is adopted to derive $\Sigma _{\mathrm{H_{2}}}$. 

The radial profile of $\Sigma _{\mathrm{H_{2}}}$ is shown as crosses in Figure \ref{FIG_radialplot}. 
$\Sigma _{\mathrm{H_{2}}}$ peaks at the galactic center ($\sim$750 M$_{\solar}$ pc$^{-2}$) and then gradually decreases to $\sim$ R/R$_{25}$ = 0.05. A small bump at a radius of R/R$_{25}\approx0.1$ ($\sim$2 kpc)  is caused by the brightest spiral arm. Beyond the spiral arm, $\Sigma _{\mathrm{H_{2}}}$ decreases to $\sim$20 M$_{\solar}$ pc$^{-2}$ at R/R$_{25}$ = 0.3.

The column density of atomic hydrogen (HI) is calculated by 
\begin{equation} 
N(\mathrm{HI}) (\mathrm{cm^{-2}})=1.823\times 10^{18}\int T_{\mathrm{b}}\mathrm{d}v,
\label{EQU_NHI}
\end{equation} 
where $T_{\mathrm{b}}$ is the brightness temperature \citep{Wil09}.
The step size between each azimuthal average of HI data is 20$\arcsec$ (320 pc), starting from 0$\arcsec$ to 1300$\arcsec$ (R/R$_{25}$ $\approx$ 0.97).
The radial profile of $\Sigma _{\mathrm{HI}}$ is shown in Figure \ref{FIG_radialplot} with a plus symbol. 
 As seen in most spiral galaxies, the distribution of HI emissions extends well beyond the molecular disk.
$\Sigma _{\mathrm{HI}}$ hits the lowest point at the galactic center and shows an upward trend until R/R$_{25}$ = 0.3.     
Then, $\Sigma _{\mathrm{HI}}$ gently declines to R/R$_{25}\approx 1$. $\Sigma _{\mathrm{HI}}$ varies within a narrow range of 1 -- 5  M$_{\solar}$ pc$^{-2}$, occupying $\sim$ 35\% of the total gas.  Hence, IC 342 is an H$_{2}$ dominant spiral galaxy, with potential for significant star formation.

\subsection{Radial Distribution of Star Formation}
\label{subsec_SFR}
The 24 $\mu$m emissions trace the dust grains heated by the UV photons from young stars and therefore serve as a tracer of obscured SFR with a timescale no longer than 10 Myr \citep{Cal05}. On the other hand, H$\alpha$ traces the ionized gas surrounding the massive stars and thereby is used as a tracer of unobscured SFR \citep{Cal05}. Hence, 24 $\mu$m + H$\alpha$ is often used to obtain the total SFR in a given region (e.g., \cite{Cal07,Ken07}). 

However, because IC 342 is located behind the galactic plane ($b$=10$\fdg$5), H$\alpha$ emissions suffer severely from dust extinction and, moreover, there is no available digital H$\alpha$ image to access. For these reasons we use an infrared image as the SFR indicator.  
The monochromatic infrared SFR is calculated as 
 \begin{equation} 
\mathrm{SFR(M_{\solar}\: yr^{-1})}=1.31\times 10^{-38}[L_{\mathrm{24\mu m}}(\mathrm{ergs\: s^{-1}})]^{0.885}
\label{EQU_SFR}
\end{equation} 
\citep{Cal07,Ca12a}. 
Equation (\ref{EQU_SFR}) is calibrated for a local SFR with a spatial scale of $\sim$ 500 pc and is valid  in the range of  $1\times 10^{40}\lesssim[L_{\mathrm{24\mu m}}]\lesssim 3\times 10^{44}$. IC 342 is within this range.
The uncertainty of the power and the coefficient are 0.02 and 15\%, respectively.

We have smoothed the 24 $\mu$m image from the resolution of $\sim$6$\arcsec$ to 20$\arcsec$ to match with the CO image. 
The 24$\mu$m image is also re-gridded to match with the CO image for further analysis. For this reason the step size and the radial range of $\Sigma _{\mathrm{SFR}}$ (triangles in Figure \ref{FIG_radialplot}) is the same as $\Sigma _{\mathrm{H_{2}}}$ (crosses). The radial distribution of $\Sigma _{\mathrm{SFR}}$ resembles that of $\Sigma _{\mathrm{H_{2}}}$ in IC 342.

The characteristic features of the radial distribution of $\Sigma _{\mathrm{H_{2}}}$, $\Sigma _{\mathrm{HI}}$, and $\Sigma _{\mathrm{SFR}}$  are summarized as follows: (a) the radial distribution of $\Sigma _{\mathrm{H_{2}}}$ and $\Sigma _{\mathrm{HI}}$ reveal that IC 342 is a molecular-dominant galaxy; (b) the molecular gas is concentrated in the center and the bright spiral arm at the radius of $\sim$2 kpc; (c) the radial distribution of $\Sigma _{\mathrm{SFR}}$ resembles that of  $\Sigma _{\mathrm{H_{2}}}$.

\section{Surface Density of H$_{2}$ Derived from $^{12}$CO} 
\label{Sec_Xcolow}
In this section, we examine the spatial distribution of molecular gas by using the Galactic CO-to-H$_{2}$ conversion factor ($X_{\mathrm{CO}}$) and a conversion factor ($X_{\mathrm{CO,v}}$) that depends on the metallicity and line intensity \citep{Nar12}.

\subsection{Surface Density of H$_{2}$ Derived with $X_{\mathrm{CO}}$}
\label{Mass12const}
First, the Galactic $X_{\mathrm{CO}}$ of 2.3 $\times$ 10$^{20}$ cm$^{-2}$ (K km s$^{-1}$)$^{-1}$ was applied to the entire $^{12}$CO map.  

The peak value of $\Sigma _{\mathrm{H_{2}}}$ is 756 M$_{\solar}$ pc$^{-2}$ at the galactic center.
Such high $\Sigma _{\mathrm{H_{2}}}$ is commonly found in the circumnuclear starburst   \citep{Ken98}. 
The 3$\sigma$ detection limit of $^{12}$CO corresponds to  $\Sigma _{\mathrm{H_{2}}}$ $\approx$ 15 M$_{\solar}$ pc$^{-2}$.
The mean $\Sigma _{\mathrm{H_{2}}}$ of IC 342 is 36 M$_{\solar}$ pc$^{-2}$, principally contributed by the galactic disk.

\subsection{Surface Density of H$_{2}$ Derived with $X_{\mathrm{CO,v}}$}
\label{Mass_Var12}
Both observations (e.g., \cite{Ari96}) and simulations (e.g., \cite{Nar12}) suggest that metallicity is a key parameter affecting $X_{\mathrm{CO}}$. In metal-poor regions, $X_{\mathrm{CO}}$ tends to be higher than the Galactic $X_{\mathrm{CO}}$ because of the photodissociation of CO. Furthermore, \citet{Nar12} show that $X_{\mathrm{CO}}$ depends on the CO intensity as well. Such dependence of $X_{\mathrm{CO}}$ on CO intensity is also found within a single galaxy \citep{Nak95}.

Therefore, we adopt the $X_{\mathrm{CO}}$--$Z{}'$--$W_{\mathrm{CO}}$ relation derived by \citet{Nar12}:
\begin{equation} 
X_{\mathrm{CO,v}}=\frac{6.75\times 10^{20}\times {\left \langle W_{\mathrm{CO}} \right \rangle}^{-0.32}}{Z{}'^{0.65}},
\label{EQU_Xco}
\end{equation} 
where  $\left \langle W_{\mathrm{CO}} \right \rangle$  is the $^{12}$CO line intensity measured in K km s$^{-1}$ and  $Z{}'$ is the metallicity divided by the solar metallicity ($Z$). 
The conversion between $Z{}'$ and the commonly used form of metallicity of  $12+\log [\mathrm{O/H}]$ is
\begin{equation} 
\log Z{}'=12+\log [\mathrm{O/H}] -8.76,
\label{EQU_Zgrad}
\end{equation} 
where the solar metallicity on a $\log$ scale is assumed to be 8.76 \citep{Kru08}.
Equation (\ref{EQU_Xco}) is available from sub-kpc scale to unresolved galaxies \citep{Nar12}.

For IC 342, as in most of galaxies, the gradient of metallicity is measured from the HII regions in various radii of the galaxy.
\citet{Pil04} derived the oxygen abundance for the HII regions observed by \citet{McC85} and fitted the gradient with the form:
\begin{equation} 
12+\log [\mathrm{O/H}] = 12+\log [\mathrm{O/H}]_{0} +C_{\mathrm{O/H}}\times (R/R_{25}),
\label{EQU_OHgrad}
\end{equation} 
where $12+\log [\mathrm{O/H}]_{0}$ is the extrapolated central oxygen abundance, and $C_{\mathrm{O/H}}$ is the slope of the oxygen abundance gradient with respect to the fractional radius $R/{R_{25}}$. The HII regions in IC 342 indicate a clear negative gradient of the abundance with increasing radius.
The value of $12+\log [\mathrm{O/H}]_{0}$ is 8.85, and $C_{\mathrm{O/H}}$ is --0.9. The uncertainty of the derived value of  $\log [\mathrm{O/H}]$ is 0.12. 
The metallicity of each individual pixel is calculated and substituted into Equation (\ref{EQU_Xco}) and  (\ref{EQU_Zgrad}) to obtain $X_{\mathrm{CO,v}}$ at each position.

$X_{\mathrm{CO,v}}$ has a range (1 -- 6) $\times$ 10$^{20}$ cm$^{-2}$ (K km s$^{-1}$)$^{-1}$.
Figure \ref{FIG_XcoXcov} displays a correlation of $\Sigma _{\mathrm{H_{2}}}$($X_{\mathrm{CO}}$) versus the ratio of $\Sigma _{\mathrm{H_{2}}}$($X_{\mathrm{CO,v}}$)/$\Sigma _{\mathrm{H_{2}}}$ ($X_{\mathrm{CO}}$) or $X_{\mathrm{CO,v}}$. The figure indicates that $X_{\mathrm{CO,v}}$ $<$ $X_{\mathrm{CO}}$ when $\Sigma _{\mathrm{H_{2}}}$($X_{\mathrm{CO}}$) $>$ 100 M$_{\solar}$ pc$^{-2}$, and vice versa. In addition, $\Sigma _{\mathrm{H_{2}}}$($X_{\mathrm{CO}}$) $\approx$ $\Sigma _{\mathrm{H_{2}}}$($X_{\mathrm{CO,v}}$) occurs at $\sim$100 M$_{\solar}$ pc$^{-2}$.    

The $X_{\mathrm{CO,v}}$ map is shown in Figure \ref{FIG_Xco}. In the galactic center, $X_{\mathrm{CO,v}}$ is as low as 10$^{20}$ cm$^{-2}$ (K km s$^{-1}$)$^{-1}$, consistent with the value measured by the high resolution C$^{18}$O, 1.3 mm dust continuum, and the virial theorem in IC 342 \citep{Mei01}. Thus the utilization of the Galactic standard $X_{\mathrm{CO}}$ leads to the overestimation of $\Sigma _{\mathrm{H_{2}}}$ at the central region. The peak $\Sigma _{\mathrm{H_{2}}}$ becomes 344 M$_{\solar}$ pc$^{-2}$ by using $X_{\mathrm{CO,v}}$, which is approximately two times lower than the mass obtained from the Galactic $X_{\mathrm{CO}}$. 
The compact regions in the  spiral arms have $X_{\mathrm{CO,v}}$ close to the Galactic $X_{\mathrm{CO}}$. The remaining diffuse parts have $X_{\mathrm{CO,v}}$ $>$ 3 $\times$ 10$^{20}$ cm$^{-2}$ (K km s$^{-1}$)$^{-1}$. Eventually, the detection limit (3 $\sigma$ detection) of $\Sigma _{\mathrm{H_{2}}}$ based on $X_{\mathrm{CO,v}}$ is about 26 M$_{\solar}$ pc$^{-2}$ and the mean $\Sigma _{\mathrm{H_{2}}}$ is approximately 51 M$_{\solar}$ pc$^{-2}$.

The radial distribution of $\Sigma _{\mathrm{H_{2}}}$ based on $X_{\mathrm{CO,v}}$  is plotted as circles in Figure \ref{FIG_radialplot}. $\Sigma _{\mathrm{HI}}$ is still negligible in the central region and the disk.


\section{Surface Density of H$_{2}$ Derived from $^{13}$CO}
\label{KS13_high}  
\begin{figure*}
\begin{center}
\FigureFile(80mm,80mm){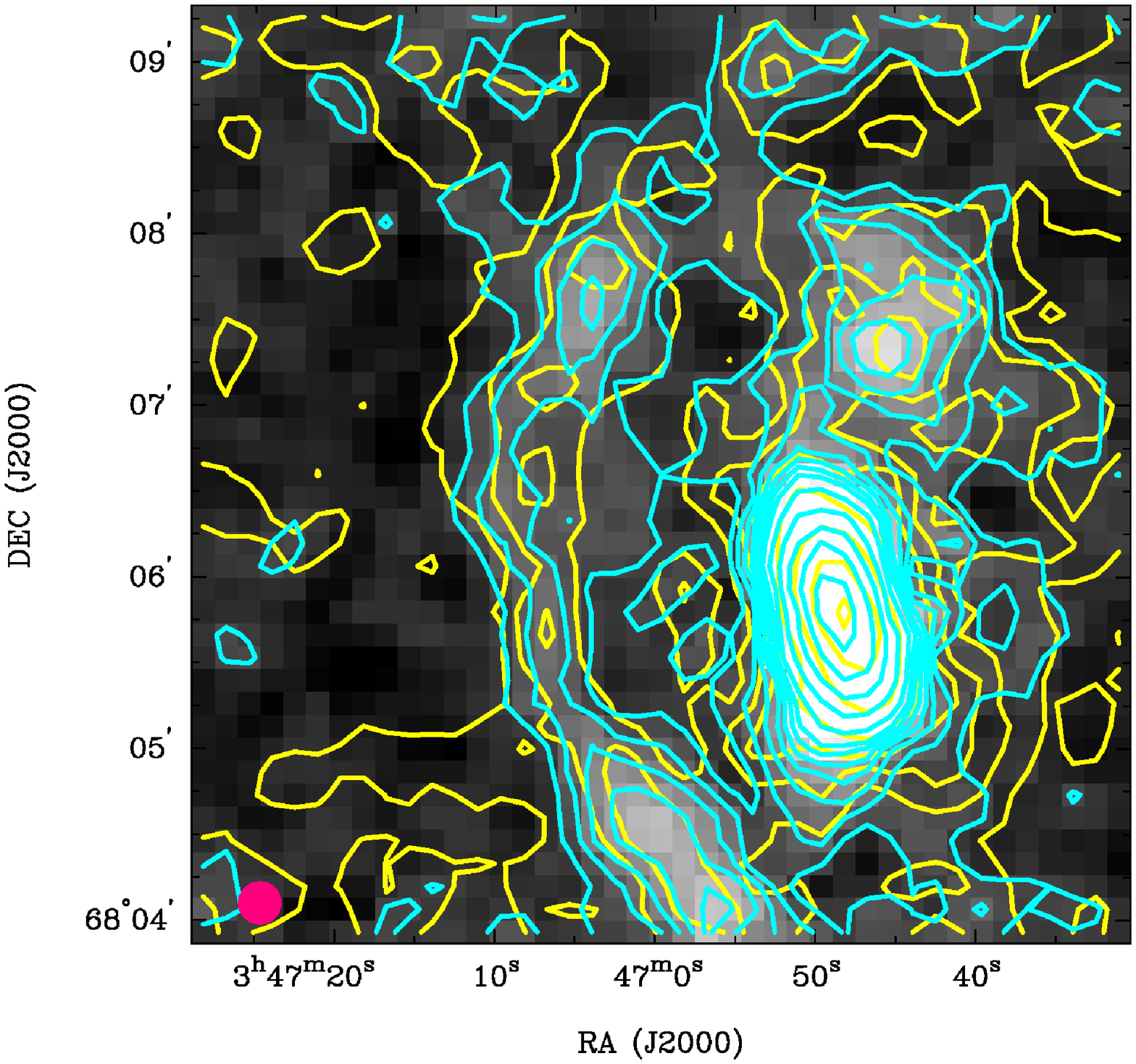}
\FigureFile(80mm,80mm){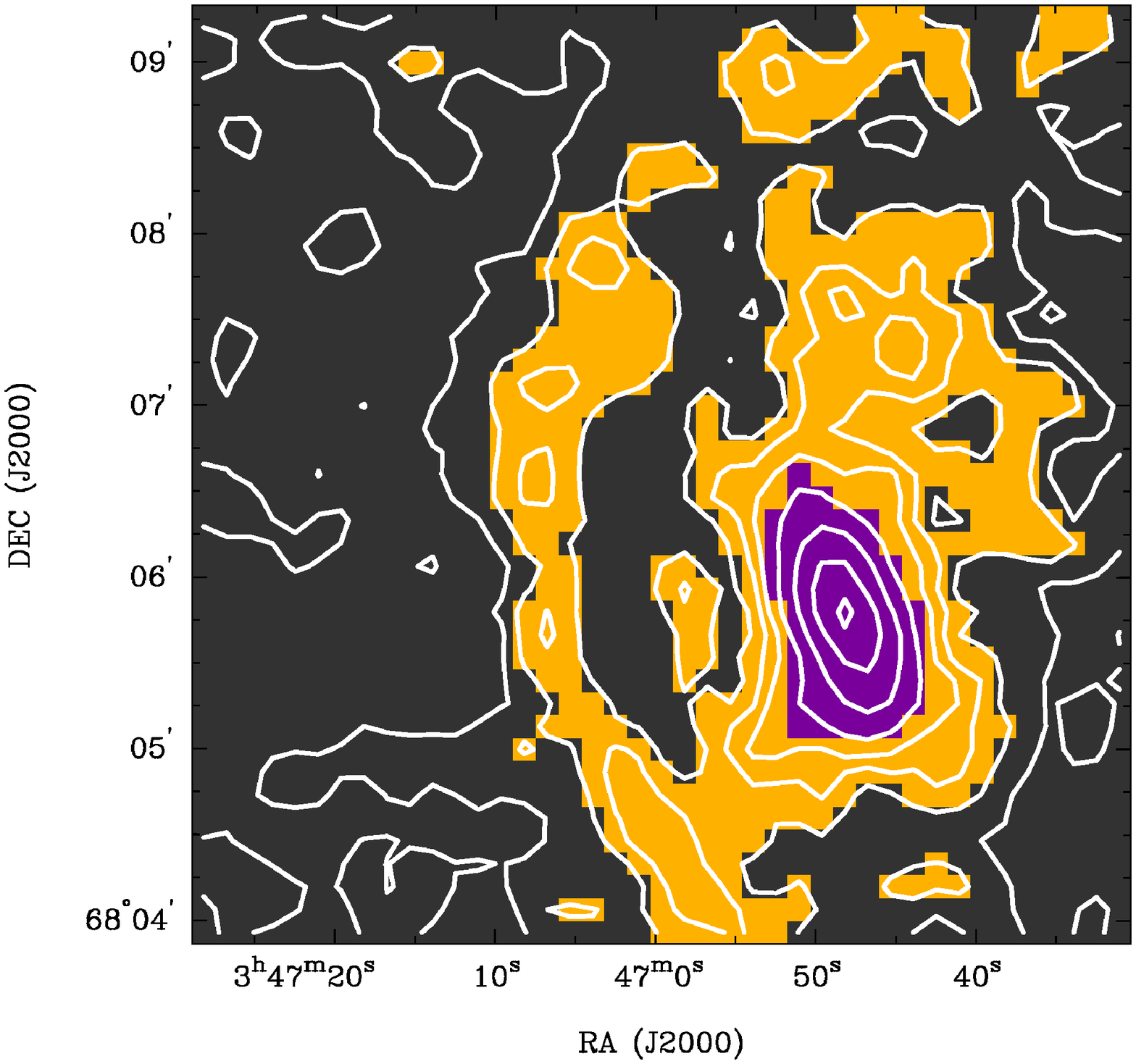}
\end{center}
\caption{
Velocity-integrated intensity maps of CO and the defined galactic features.
Left: The integrated maps of two CO lines. 
Cyan and yellow contours denote the $^{12}$CO and $^{13}$CO velocity-integrated intensity maps, respectively. $^{12}$CO is shown in Figure \ref{FIG_ico}. The beam size is overlaid in the bottom left of the figure.  The grey scale is also the $^{12}$CO map.
The contours of the $^{12}$CO map are 10 (6.5 $\sigma$), 15, 20, 25, 30, 35, 40, 50, 70, 100, 140, and 190 K km s$^{-1}$.
The contour levels of the $^{13}$CO map are 1 (2.5 $\sigma$), 2, 3, 4, 6, 10, 16, and 22 K km s$^{-1}$. Right: the defined galactic features. The white contours denote the $^{13}$CO in the left panel with the same steps. The purple area corresponds to the pixels categorized as the galactic center, and the orange area tags the pixels belonging to the defined spiral arm.}
\label{FIG_structure}
\end{figure*}

\begin{figure}
\begin{center}
\FigureFile(90mm,150mm){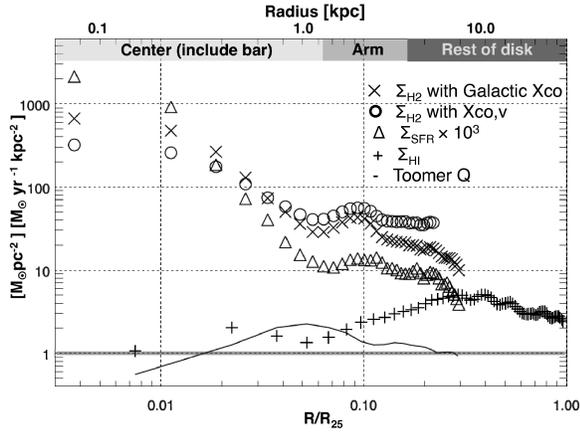}
\end{center}
\caption{Radial distribution of $\Sigma _{\mathrm{H_{2}}}$ (crosses and circles), $\Sigma _{\mathrm{HI}}$ (pluses) and $\Sigma _{\mathrm{SFR}}$ (triangles). The radial step of $\Sigma _{\mathrm{H_{2}}}$ and $\Sigma _{\mathrm{SFR}}$ is 10$\arcsec$. The radial profile of $\Sigma _{\mathrm{HI}}$ is sampled every 20$\arcsec$. The solid black line represents the Toomre Q parameter (\S\ref{SF_IC342}). The grey solid line marks an intensity of one for comparison with Q.}
\label{FIG_radialplot}
\end{figure}

\begin{figure}
\begin{center}
\FigureFile(80mm,80mm){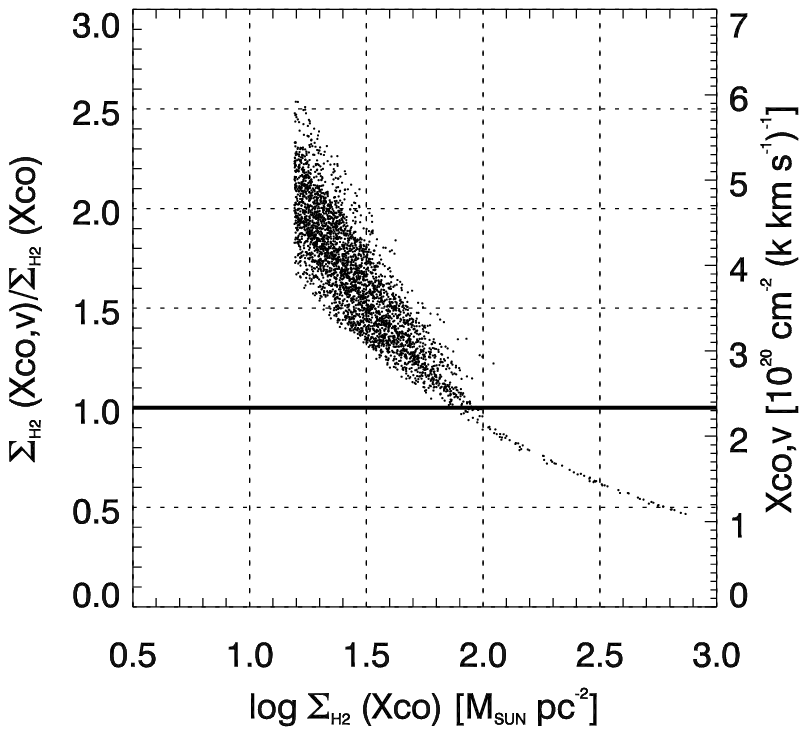}
\end{center}
\caption{$\Sigma _{\mathrm{H_{2}}}$($X_{\mathrm{CO}}$) versus the ratio of $\Sigma _{\mathrm{H_{2}}}$($X_{\mathrm{CO,v}}$)/$\Sigma _{\mathrm{H_{2}}}$ ($X_{\mathrm{CO}}$) and $X_{\mathrm{CO,v}}$. The horizontal line marks where $\Sigma _{\mathrm{H_{2}}}$($X_{\mathrm{CO,v}}$)/$\Sigma _{\mathrm{H_{2}}}$($X_{\mathrm{CO}}$) $=$ 1.}
\label{FIG_XcoXcov}
\end{figure}

\begin{figure}
\begin{center}
\FigureFile(80mm,80mm){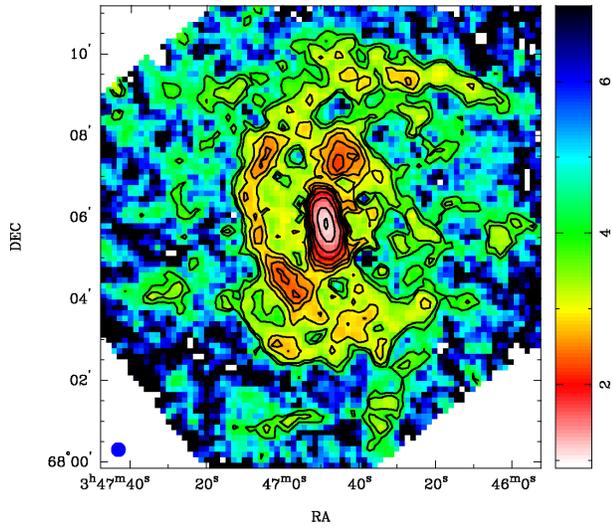}
\end{center}
\caption{The $X_{\mathrm{CO,v}}$ map of IC 342. The right color bar of $X_{\mathrm{CO,v}}$ is in unit of 10$^{20}$ cm$^{-2}$ (K km s$^{-1}$)$^{-1}$. The beam size of 20$\arcsec$ is indicated in the lower left.}
\label{FIG_Xco}
\end{figure}

The velocity-integrated intensity map of $^{13}$CO acquired from \citet{Hir10} is shown in the left panel of Figure \ref{FIG_structure} with yellow contours. 
The $^{13}$CO to $^{12}$CO ratio is 0.05 -- 0.2 \citep{Hir10}.
Owing to the low intensity of $^{13}$CO, only the pixels with significant detection are considered. 
The central $\sim$1$\arcmin$ corresponding to the galactic nucleus found in the high-resolution map \citep{Mei05} is defined as {\it galactic center}. This region has at least 15 $\sigma$ detections. 
The central region is colored purple in the right panel of Figure \ref{FIG_structure}. Apart from the galactic center, the pixels with $\geq$ 5 $\sigma$ detection are taken as well. The criterion chiefly captures the primary spiral arm and the bar end. We refer to this region as the {\it spiral arm}. This region is colored orange in Figure \ref{FIG_structure}.
   
Provided that $^{13}$CO is in local thermal equilibrium (LTE) and optically thin, the column density of H$_{2}$ is computed by  
\begin{eqnarray}
N(\mathrm{H_{2}})\mathrm{_{^{13}CO}} &=& 2.41\times 10^{14}\times\frac{\tau \left ( ^{13}\mathrm{CO} \right )}{1-\mathrm{e}^{-\tau (^{13}\mathrm{CO})}} \nonumber \\
&\times & \mathrm{\frac{[H_{2}]}{[^{13}CO]}}\times \frac{I_{\mathrm{^{13}CO}}}{1-e^{-5.29/T\mathrm{_{ex}}}}, 
\label{EQU_N13CO} 
\end{eqnarray} 
where $\tau$ is the optical depth of the $^{13}$CO, [H$_{2}$]/[$^{13}$CO] is the inverse of the $^{13}$CO abundance, $I_{\mathrm{^{13}CO}}$ is the intensity of $^{13}$CO in K km s$^{-1}$, and $T_{\mathrm{ex}}$ is the excitation temperature in Kelvin \citep{Wil09}. As shown in Equation (\ref{EQU_N13CO}),  prior knowledge of the optical depth,  $^{13}$CO abundance, and gas (excitation) temperature are required.

The optical depth is estimated by 
\begin{equation} 
\tau\left ( \mathrm{^{13}CO} \right )\approx I_{\mathrm{^{13}CO}}/I_{\mathrm{^{12}CO}},
\end{equation} 
in which $^{12}$CO is assumed to be optically thick and $^{13}$CO is optically thin \citep{Pag01}.

The isotopic abundance ratio [$^{12}$C]/[$^{13}$C] is assumed to be $\approx$ 30-50 from the measurements for the galactic center of IC 342 \citep{Hen93,Hen98}. Given that [$^{12}$C]/[$^{13}$C] $=$  [$^{12}$CO]/[$^{13}$CO] (e.g., \cite{Esp10}) and  [$^{12}$CO]/[${\mathrm{H_{2}}}$] is approximately (5 -- 8) $\times$ 10$^{-5}$, we adopt a constant [$^{13}$CO]/[${\mathrm{H_{2}}}$] of 1.5 $\times$ 10$^{-6}$ in all regions concerned. The value is consistent with the mean $^{13}$CO abundance observed in the Galactic star forming clouds \citep{Pin08,Pin10}.

The gas temperature is also required to obtain the column density of H$_{2}$. 
The assumption that \textit{the gas temperature ($T _{\mathrm{g}}$) is equal to the dust temperature ($T _{\mathrm{d}}$)} is made. Then, the dust/gas temperature can be yielded by the infrared SED.

\begin{figure*}[]
\begin{center}
\FigureFile(50mm,50mm){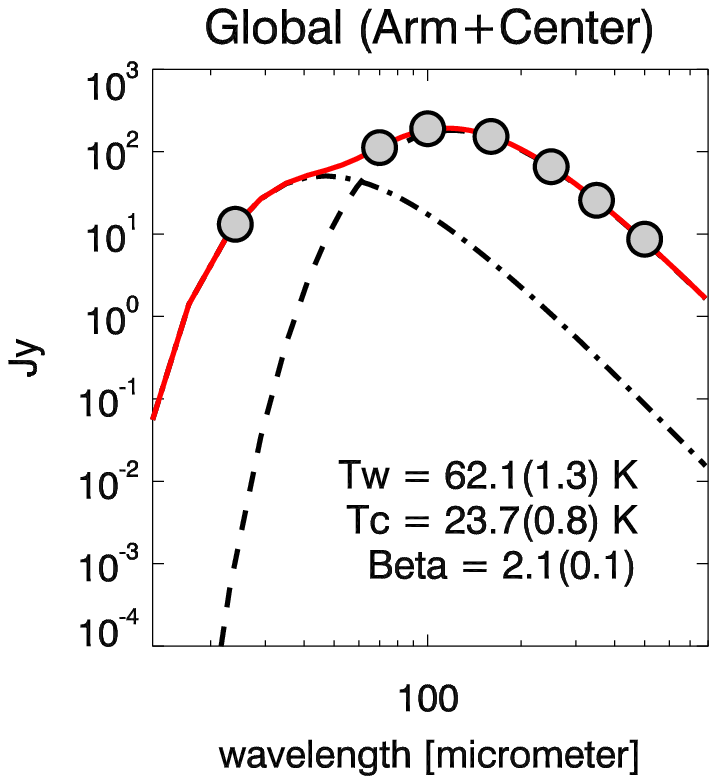}
\FigureFile(50mm,50mm){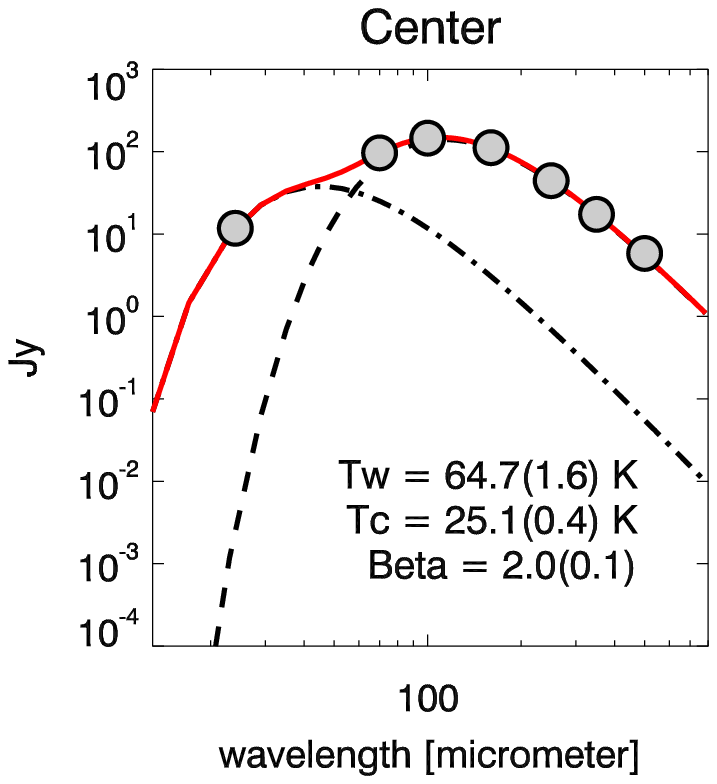}
\FigureFile(50mm,50mm){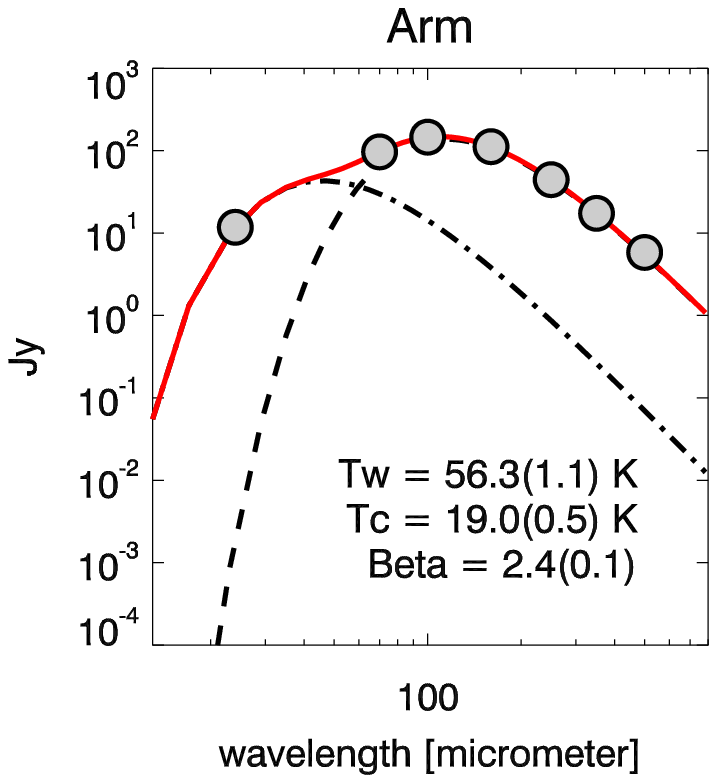}
\end{center}
\caption{Infrared SEDs in log--log scale. The global (arm + center) SED is shown in the left, the central and arm SEDs are displayed in the middle and on the right, respectively. The circles are the observed data points. The dash-dotted and dashed curves are the derived Planck distribution of the warm and cold components, respectively. The red curve is the sum of two components. The derived $T_{\mathrm{w}}$, $T_{\mathrm{c}}$, and $\beta_{\mathrm{c}}$ are presented in each panel. The error of each parameter is shown in parentheses.}
\label{FIG_T}
\end{figure*}

The SED is constructed with the 24$\mu$m observations by Spitzer (\S\ref{sec_data_sp}) and the 70 -- 500$\mu$m observations by Herschel (\S\ref{sec_data_her}).
All images are converted to janskys and then convolved to the lowest resolution among the images, i.e.,  FWHM = 37$\arcsec$ of the Herschel 500 $\mu$m map.
The method of convolution kernels between Spitzer and Herschel images is described in \citet{Ani11}. We used the kernels and the IDL procedure ${\tt conv\_image}$ \citep{Gor08} to convolve the images.
Then, a two-component Planck distribution
\begin{equation} 
D=C_{\mathrm{w}}\; \left [ \lambda ^{-2}\; B_{\nu}\left ( T_{\mathrm{w}} \right )  \right ]+C_{\mathrm{c}}\; \left [ \lambda ^{-\beta_{\mathrm{c}}}\; B_{\nu}\left ( T_{\mathrm{c}} \right )  \right ]
\label{EQU_2compt}
\end{equation}
was assumed as the shape of SED, where $B_{\nu}$ is a Planck function, $T_{\mathrm{w}}$ and $T_{\mathrm{c}}$ are the temperatures of the warm and cold components, respectively, and $C_{\mathrm{w}}$ and $C_{\mathrm{c}}$ are the scaling constants of each component \citep{Gal12}. Moreover, the Planck functions were modified by the dust emissivity index $\beta$. For the warm component, the warm dust emissivity index ($\beta_{\mathrm{w}}$) was fixed at two as suggested by \citet{Li01}. On the other hand, the cold dust emissivity index ($\beta_{\mathrm{c}}$) is a free parameters to be fit. $\beta_{\mathrm{c}}$ is in the range of 1 -- 3, so the program is forced to find the best value of $\beta_{\mathrm{c}}$ in this range.
To summarize, there are five free parameters to fit: $C_{\mathrm{w}}$, $C_{\mathrm{c}}$, $T_{\mathrm{w}}$, $T_{\mathrm{c}}$, and $\beta_{\mathrm{c}}$.
Finally, the IDL function ${\tt MPCURVEFIT}$ \citep{Mar09} is used to carry out a Levenberg--Marquardt least--squares fit of the data. ${\tt MPCURVEFIT}$ requires a user-supplied function for fitting. In this case, Equation (\ref{EQU_2compt}) is the function employed.
Since the images have a final resolution significantly lower than our CO maps, SEDs for the global (center + arm), galactic center and arm are obtained with the total flux in each defined region in Figure \ref{FIG_structure}, instead of a pixel-by-pixel fitting. The SEDs are displayed in Figure \ref{FIG_T}. The derived $\beta_{\mathrm{c}}$ is in the typical range of 2.0 -- 2.4. The galactic center has a higher temperature in both warm and cold components (warm: 64.7 $\pm$ 1.6 K, cold: 25.1 $\pm$ 0.4 K) than those of the spiral arm (warm: 56.3 $\pm$ 1.1 K, cold: 19.0 $\pm$ 0.5 K). The global SED represents the average of the center and disk with $T_{\mathrm{w}}$ $=$ 62.1 $\pm$ 1.3 K and $T_{\mathrm{c}}$ $=$ 23.7 $\pm$ 0.8 K. 
\citet{Gal12} derived $19\, \mathrm{K}< T_{\mathrm{c}}< 25\,\mathrm{K}$ and  $55\, \mathrm{K}< T_{\mathrm{c}}< 63\,\mathrm{K}$ in eleven nearby galaxies with similar methods and similar datasets. The temperatures of IC  342 are compatible with other galaxies.

The cold dust component is associated with the dust, molecular, and atomic hydrogen clouds mostly heated by the interstellar radiation field (ISRF). On the other hand, the warm dust component is heated by the OB stars in HII regions \citep{Cox89,Pop02, Ber10}.
Accordingly, $T_{\mathrm{c}}$ ($\sim$25 K for the center and $\sim$19 K for the arm) is substituted into Equation (\ref{EQU_N13CO}) to derive $\Sigma _{\mathrm{H_{2}}}$ based on $^{13}$CO. The resulting $\Sigma _{\mathrm{H_{2}}}$ lies within 30 -- 400 $M_{\solar}$ pc$^{-2}$. The mean $\Sigma _{\mathrm{H_{2}}}$ is 90 $M_{\solar}$ pc$^{-2}$. The mean $\Sigma _{\mathrm{H_{2}}}$ obtained by $^{13}$CO is significantly larger than those from $^{12}$CO (\S\ref{Sec_Xcolow}) because the observations and the selection of available pixels are biased to the CO bright region as a result of the weak intensity of $^{13}$CO.

\begin{figure}
\begin{center}
\FigureFile(80mm,80mm){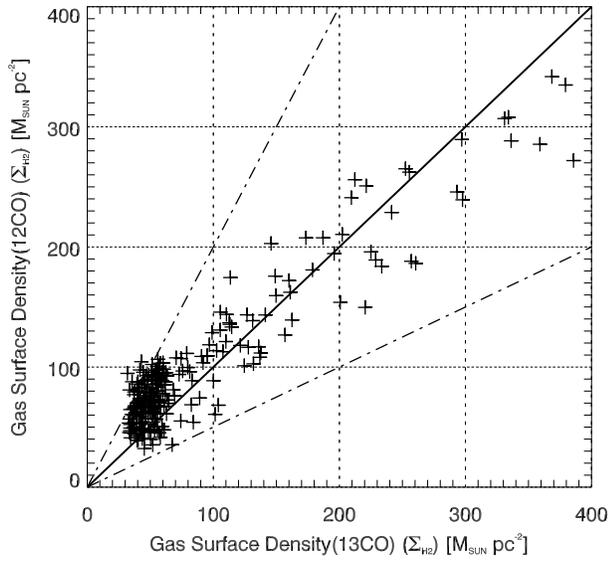}
\end{center}
\caption{$\Sigma _{\mathrm{H_{2}}}$ ($^{13}$CO) versus $\Sigma _{\mathrm{H_{2}}}$ ($^{12}$CO, $X_{\mathrm{CO,v}}$). The solid black line indicates the slope of unity. The upper and lower dash-dotted lines represent the slopes of 2.0 and 0.5, respectively.
}
\label{FIG_MassMass}
\end{figure}

$\Sigma _{\mathrm{H_{2}}}$ obtained from $^{12}$CO($X_{\mathrm{CO,v}}$) and $^{13}$CO are compared in Figure \ref{FIG_MassMass}. Only the corresponding pixels in the $^{12}$CO map with significant $^{13}$CO detection (Figure \ref{FIG_structure} right panel) are selected for comparison. 
$\Sigma _{\mathrm{H_{2}}}$ derived from both lines are consistent where $\Sigma _{\mathrm{H_{2}}}$ $>$ 100 M$_{\solar}$ pc$^{-2}$. Such high $\Sigma _{\mathrm{H_{2}}}$ is dominated by the central region of the galaxy.

However, $\Sigma _{\mathrm{H_{2}}}$ acquired from $^{13}$CO is smaller than that from $^{12}$CO at $\Sigma _{\mathrm{H_{2}}}$ $\lesssim$ 100 M$_{\solar}$ pc$^{-2}$.
The deviation of $\Sigma _{\mathrm{H_{2}}}$ between two lines can be attributed to the overestimation of the  $^{13}$CO abundance by adopting the central value even outside the center. The uncertainty of the $^{13}$CO abundance will be discussed in the next section (\S\ref{uncertainty_Cabundance}).
Nevertheless, we should note that in the spiral arm, $^{12}$CO and $^{13}$CO peaks do not coincide with each other (see Figure \ref{FIG_structure}). Therefore, it is not necessary to expect a consistent result from these lines.

\section{Uncertainties of $\Sigma _{\mathrm{H_{2}}}$}
\label{UncerMass}
\subsection{Uncertainty of Radial Oxygen Abundance}  
\label{uncertainty_Oxyabundance} 
To obtain the $\Sigma _{\mathrm{H_{2}}}$ map based on $^{12}$CO, the gradient of oxygen abundance is required.  
The emission lines of HII regions spanning a large fraction of the galactic disk are needed to construct the metallicity gradient of a galaxy.
The relevant studies have demonstrated that the metallicity (12+[O/H]) is a function of the galactocentric radius in spiral galaxies, although the slopes can vary between positive and negative \citep{Wer11}. For IC 342, only five HII regions were observed to infer the metallicity gradient \citep{McC85}. It seems that the gradient is affected by the number of HII regions observed.
\citet{Mou10}, however, derive the gradient of metallicity with eight HII regions for NGC 6946, classified as SAB(rs)cd as IC 342, and the result is consistent with the results derived with more than 200 HII regions in the galaxy \citep{Ced12}.
The comparison demonstrates that the variation of oxygen abundance is on a global scale and the local fluctuation is not significant. The uncertainty of $X_{\mathrm{CO,v}}$ attributed to the oxygen abundance alone is $\sim$40\%. The uncertainty is dominated by the scatter of the measured oxygen abundance from the general trend at each position.

\subsection{Uncertainty of $^{13}$CO abundance}
\label{uncertainty_Cabundance}
The isotopic ratio [$^{12}$C]/[$^{13}$C] grows with radius in our Galaxy \citep{Mil05}. 
However, a constant $^{13}$CO abundance measured for the center is adopted for the entire region  in this work because of the lack of abundance ratio along the galactocentric radius of IC 342.

$^{13}$C is produced during the CNO cycles of stars. In an active star forming region, e.g., galactic center, $^{13}$C is produced faster, influencing the interstellar isotope abundances.
There are two subsequent mechanisms determining the $^{13}$CO abundance: fractionation reaction and photodissociation.
The fractionation reaction $\mathrm{^{13}C^{+}+^{12}CO \to ^{13}CO+^{12}C^{+}}$ enriches $^{13}$CO but requires relatively low temperature ($<$ 35K) \citep{Wat76} and UV photons to provide the ionized carbons, preferentially happening near the UV-exposed edges of GMCs.
However, UV photons cause the photodissociation of both $^{12}$CO and $^{13}$CO as well ($\mathrm{CO}+h\nu \rightarrow \mathrm{C+O}$). 
Because self-shielding is more effective for H$_{2}$ than CO lines,  photodissociation can reduce the $^{12}$CO and $^{13}$CO abundance relative to H$_{2}$.
In addition, because $^{12}$CO has more effective self-shielding and larger abundance, the photodissociation destroys $^{13}$CO faster than $^{12}$CO.  
Combining both phenomena, the abundance of $^{13}$CO depends on the relative rate of the fractionation and photodissociation.
The combined effect can be small, and these phenomena compensate for one another \citep{Kee98}, or leads to the underestimation of $\Sigma_{\mathrm{H_{2}}}$ by 30\%--50\% \citep{Gol08,Wol10}. 
Providing that  [$^{12}$C]/[$^{13}$C] (or [$^{12}$CO]/[$^{13}$CO] as we have assumed) has the same gradient as our Galaxy  \citep{Mil05}, the isotopic ratio then  increases from $\sim$40 at the galactic center to $\sim$50 at the spiral arm at 2 kpc. This deviation in the isotopic ratio is smaller than the uncertainties suggested by the above chemical processes.   
We have insufficient information to quantify the actual error. A maximum error of 50\% contributed from the uncertainty of $^{13}$CO abundance is quoted in this work, but such large uncertainty is unlikely to be true if the abundance gradient is similar to that of the Milky Way. 
\subsection{Uncertainty of Gas/Dust Temperatures}
 \label{uncertainty_temperature}
The interaction and energetics of gas and dust are complex.
In terms of simulations, \citet{Yao06} made starburst models to correlate the observed FIR/submm/mm properties of gas and the star formation history of the starburst galaxy. The model predicts that $T _{\mathrm{g}}$ is close to $T _{\mathrm{d}}$ in high-column-density regions.
\citet{Gol01} use LVGs model to calculate the radiative transfer of the cooling and heating of gas and dust. The result shows that in the high-volume-density region with  $n\;\geq$ 10$^{4.5}$  cm$^{-3}$ the gas and dust grains become well coupled and $T _{\mathrm{g}}$ $\approx$ $T _{\mathrm{d}}$. Recent work has also suggested a similar threshold of 10$^{4}$ cm$^{-3}$\citep{Nar11,Mar12} for good coupling  between $T _{\mathrm{g}}$ and $T _{\mathrm{d}}$.
Even though the critical density of $^{13}$CO is $\sim$10$^{3}$ cm$^{-3}$, tracing the gas with density $>$ 10$^{3}$ cm$^{-3}$, $^{13}$CO is useful to locate the dense cores embedded in GMCs \citep{Mcq02}, which have a typical density of $\sim$10$^{4-7}$ cm$^{-3}$ (e.g., \cite{Ber96,Pir03}). Therefore, it is not surprising to see that $^{13}$CO coexists with 24 $\mu$m emissions in both the spiral arm and the center of IC 342 \citep{Hir10}. Hence, the assumption of $T _{\mathrm{g}}$ $\approx$ $T _{\mathrm{d}}$ is rather fair for the molecular gas traced by $^{13}$CO in terms of gas density.

The inferred dust/gas temperature from SED fitting is $\sim$ 20 -- 25 K in the infrared bright galaxy IC 342. Although the temperature is consistent with the temperature of the cloud complexes in other infrared bright galaxies (e.g., \cite{Reb12}), the Galactic GMCs have typical temperatures of $\sim$ 10 K, which is approximately two times lower than that of these infrared bright galaxies. 
Nonetheless, it is unclear whether the external galaxies, especially the infrared bright galaxies, hold gas properties similar to those in our Milky Way (e.g., \cite{Hen98}).

\section{K--S Law in IC 342}
\label{KS} 
Because $^{12}$CO is detected throughout the entire galaxy (Figure \ref{FIG_ico}), the map is first used to perform the K--S plot by using the value of $\Sigma _{\mathrm{H_{2}}}$ based on $X_{\mathrm{CO}}$ and $X_{\mathrm{CO,v}}$. The fitting of the K--S plot is dominated by the area with $\Sigma _{\mathrm{H_{2}}}$ $\ll$ 100 M$_{\solar}$ pc$^{-2}$ because of the bulk of data points in this range. We therefore refer to the results as {\it the K--S law in the low-$\Sigma _{\mathrm{H_{2}}}$ region}. On the contrary, considering the fact that $^{13}$CO was only observed toward the CO bright regions, the K--S plots made with $^{13}$CO and the corresponding $^{12}$CO pixels are referred to as {\it the K--S law in the high-$\Sigma _{\mathrm{H_{2}}}$ region}.   

Since the fitting procedures can alter $N$ of K--S law (\cite{Ca12b} and references therein), the routines we employed are described below.
The ${\tt IDL}$ routine ${\tt MPFIT}$ \citep{Mar09} is used to implement all the fitting of  the K-S plots in this work. 
Similar as the ${\tt MPCURVEFIT}$ in \S\ref{KS13_high}, ${\tt MPFIT}$ allows a user-defined function, then performs Levenberg-Marquardt least-squares minimization for the dataset. 
We fitted the data with a single power law, i.e., $y=Nx+A$, where  $y=\log _{10}(\Sigma _{\mathrm{SFR}})$, $ x=\log _{10}(\Sigma _{\mathrm{H_{2}}})$, $N$ is the slope of the K--S plot, and $A$ is a constant. 
Because  both $\Sigma _{\mathrm{H_{2}}}$ and $\Sigma _{\mathrm{SFR}}$ contain measured uncertainties, a model function ${\tt LINFITEX}$ from the ${\tt MPFIT}$ library is utilized. 
${\tt LINFITEX}$ is developed to handle a linear fitting to the data with uncertainties in both directions, following the methodology of \emph{ Numerical Recipes}.
\subsection{K--S Law at the Low-$\Sigma _{\mathrm{H_{2}}}$ Regions}
\label{KSlowSig}
\subsubsection{K--S Plot with  $X_{\mathrm{CO}}$}
\label{LargeConstant} 
\begin{figure*}
\begin{center}
\FigureFile(80mm,80mm){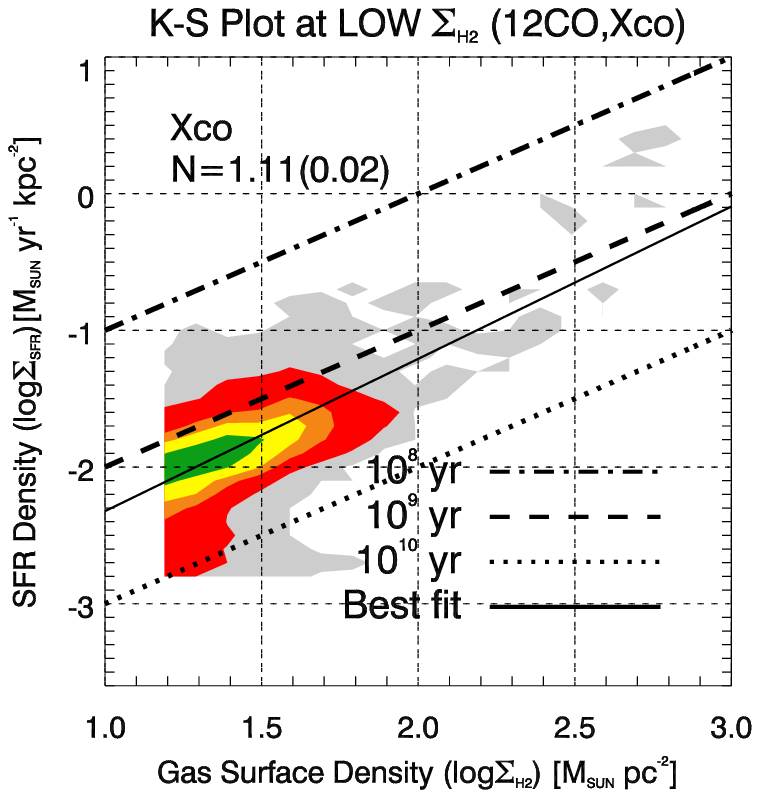}
\FigureFile(80mm,80mm){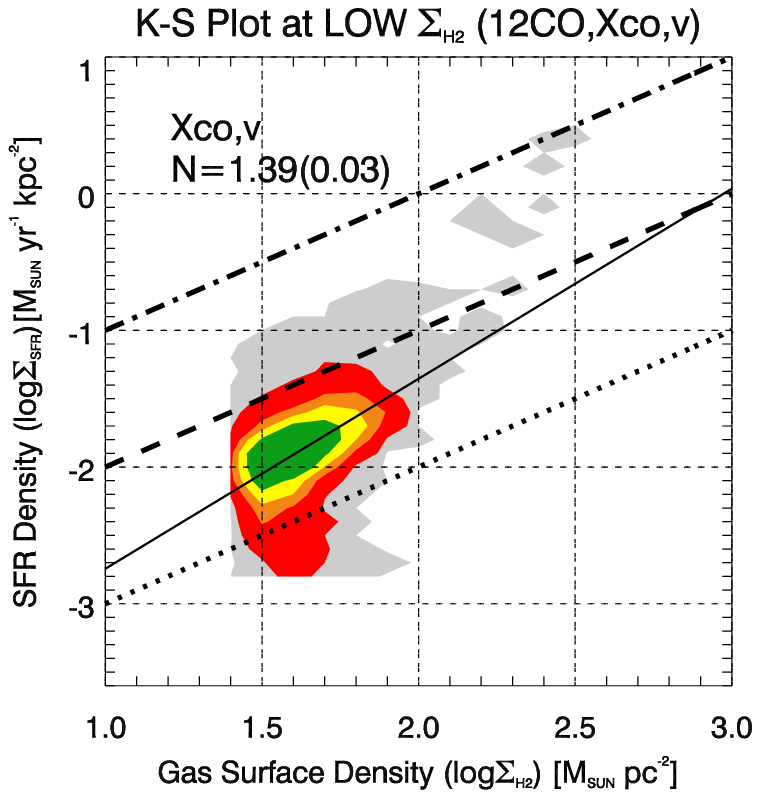}
\end{center}
\caption{
K--S plots in low-$\Sigma _{\mathrm{H_{2}}}$ region derived from $^{12}$CO. Left:  K--S plot with $X_{\mathrm{CO}}$. Grey, red, orange, yellow, and green correspond to the contours of 1, 10, 40, 65, and 100 data points per 0.1 dex-wide cell of both $\Sigma _{\mathrm{H_{2}}}$ and $\Sigma _{\mathrm{SFR}}$ axes, respectively. Fifty percent of the data points are located inside the yellow region. The best fit has a slope of 1.11 $\pm$ 0.02 shown as a solid black line. The three parallel lines denote gas depletion time of 10$^{8}$, 10$^{9}$, and 10$^{10}$ yr, corresponding to star formation efficiency of 10$^{-8}$, 10$^{-9}$, and 10$^{-10}$ yr$^{-1}$. Right: K--S plot with $X_{\mathrm{CO,v}}$. The colors of the contours and the lines are the same as in the left panel. Fifty percent of the data points are located inside the green region. The slope of the best-fit line is 1.39 $\pm$ 0.03.
}
\label{FIG_LargeSpatial_KS}
\end{figure*}

The K--S law is examined with $\Sigma _{\mathrm{H_{2}}}$ derived with the Galactic standard $X_{\mathrm{CO}}$ in \S\ref{Mass12const} and the SFR from Equation (\ref{EQU_SFR}) (The 24 $\mu$m image has been re-gridded to the same spacing and the resolution as the $^{12}$CO map in \S\ref{subsec_SFR}). 
The K--S plot is shown in the left panel of Figure \ref{FIG_LargeSpatial_KS}. 
Gray, red, orange, yellow, and green indicate the contours of 1, 10, 40, 65, and 100 data points per 0.1 dex-wide cell of both $\Sigma _{\mathrm{H_{2}}}$ and $\Sigma _{\mathrm{SFR}}$, respectively.
Fifty percent of the data points are inside the yellow area with $\Sigma _{\mathrm{H_{2}}}$ $\approx$ 15 (sensitivity limit) -- 40  M$_{\solar}$ pc$^{-2}$.  
Thus the result is controlled by the low $\Sigma _{\mathrm{H_{2}}}$ region. The slope of the K--S plot ($N$) is 1.11 $\pm$ 0.02.   

\citet{Big08} created  K--S plots for seven nearby spiral galaxies by the same method but with $^{12}$CO (2--1), assuming a constant ratio of $^{12}$CO (2--1)/(1--0). 
$\Sigma _{\mathrm{H_{2}}}$ in their samples mostly lie in the range of 3--50 M$_{\solar}$ pc$^{-2}$. The lower end of $\Sigma _{\mathrm{H_{2}}}$ in \citet{Big08} is smaller than our estimate as a result of better sensitivity.
The samples in \citet{Big08} have a physical resolution of 750 pc, approximately two times larger than in this study.
Previous works show that the physical resolution can affect  $N$ of the K--S law owing to the evolution of GMCs. 
For a small scale ($\sim$ 60 -- 100 pc),  discrete star forming systems have different evolutionary stages. For example, the youngest GMCs have no star formation but large amounts of molecular gas (large $\Sigma _{\mathrm{H_{2}}}$, low $\Sigma _{\mathrm{SFR}}$), whereas old GMCs have the opposite characteristics (small $\Sigma _{\mathrm{H_{2}}}$, large $\Sigma _{\mathrm{SFR}}$). These populations then scatter the K--S relation \citep{Ono10}.      
Our resolution of 320 pc and the 750 pc of \citet{Big08} are considerably larger than the physical sizes of the GMCs; therefore, the abovementioned situation does not exist. Instead, the K--S law represents a time-averaged relation, averaging many elements with various evolutionary stages, and thus the scaling relation of $\Sigma _{\mathrm{H_{2}}}$ and $\Sigma _{\mathrm{SFR}}$ can be maintained \citep{Sch10,Fek11}.

$N$ of 1.11 $\pm$ 0.02 in IC 342 is comparable to the average slope of 1.0 $\pm$ 0.2 among the galaxies in \citet{Big08}. 
 $N$ of unity implies that the gas depletion time ($\tau _{\mathrm{dep}}$) and star formation efficiency (SFE) are independent of $\Sigma _{\mathrm{H_{2}}}$.
The $\tau _{\mathrm{dep}}$ of IC 342 is about 2 $\times$ 10$^{9}$ yr, or SFE $\sim$ 5 $\times$ 10$^{-10}$ yr$^{-1}$ (Figure \ref{FIG_LargeSpatial_KS} left panel).

In conclusion, on the basis of  the comparison with the larger sample in \citet{Big08}, we suggest that the relation of the gas and star formation properties in IC 342 are not considerably  different from those of other galaxies.
However, we should note that in addition to the influence from the intrinsic properties of GMCs \citep{Ono10,Fek11}, \citet{Ca12b} suggest that the configuration of GMCs and stars can affect  $N$ of star formation law when observations are made with approximately sub-kiloparsec resolution.
Under this physical resolution, the distribution of molecular gas is spatially resolved, but individual GMCs are not. 
As a result, the observed $N$ is a convolution of the intrinsic star formation mode, cloud mass spectrum in the sampled area, volume filling factor, and region size (physical resolution) \citep{Ca12b}.
These parameters may vary among individual galaxies. 
In general, $N$ will decrease when the physical resolution increases, although we did not see the significant change among IC 342 (resolution of 320 pc) and the samples of \citet{Big08}  (resolution of 750 pc).

\subsubsection{K--S Plot with $X_{\mathrm{CO,v}}$}
\label{sec_LargeSpatialKSXvar}
We created a K--S plot using the same method as in \S\ref{LargeConstant} but with $X_{\mathrm{CO,v}}$ instead. The result is displayed in the right panel of Figure \ref{FIG_LargeSpatial_KS}. The color scale is the same as in \S\ref{LargeConstant}.
Fifty percent of the data points are located inside the green region ranging from 25 -- 60 M$_{\solar}$ pc$^{-2}$. 
The figure shows that the relation between $\Sigma _{\mathrm{H_{2}}}$ and $\Sigma _{\mathrm{SFR}}$ is steepened. The derived $N$ is 1.39 $\pm$0.03.   
Because of the superlinear relation, the $\tau _{\mathrm{dep}}$ and the SFE are no longer constant. Specifically,  $\tau _{\mathrm{dep}}$ declines with increasing $\Sigma _{\mathrm{H_{2}}}$, namely, the SFE rises when $\Sigma _{\mathrm{H_{2}}}$ increases. 

Because the fitting is dominated by the bulk of the data points at low-$\Sigma _{\mathrm{H_{2}}}$ regions (around 25 -- 60 M$_{\solar}$ pc$^{-2}$), the high-surface-density regions with  $\Sigma _{\mathrm{H_{2}}}$ $>$ 100 M$_{\solar}$ pc$^{-2}$ are located beyond the best fit, suggesting a steeper trend. 

\subsection{K--S Law at High-$\Sigma _{\mathrm{H_{2}}}$ Regions} 
\label{hiSig_KS} 

In this section we study the K--S law in the area with high $\Sigma _{\mathrm{H_{2}}}$.
$^{13}$CO map can be used as a filter to select the region with high  $\Sigma _{\mathrm{H_{2}}}$ because (1) $^{13}$CO is a weak line, the location where $^{13}$CO can be significantly detected is biased to high-$\Sigma _{\mathrm{H_{2}}}$ regions, and (2) owing to its higher critical density, $^{13}$CO is one of the transitions available for locating the dense cores in GMCs. These dense cores have higher column density ($\Sigma _{\mathrm{H_{2}}}$) and volume density prepared for future star formation. 

In the right panel of Figure \ref{FIG_structure}, the pixels with $^{13}$CO detection greater than 5$\sigma$ are masked with color. These pixels are used to derive a K--S plot based on $^{13}$CO.

\subsubsection{K--S plot Based on $^{13}$CO}
\label{secKS13} 
\begin{figure}
 \begin{center}
  \includegraphics[width=0.44\textwidth]{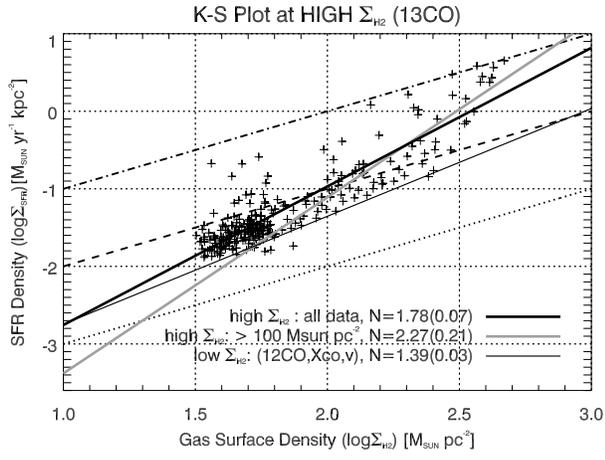}
  \caption{K--S plot in high-$\Sigma _{\mathrm{H_{2}}}$ region derived from $^{13}$CO. The best fit with all the data points is shown as a thick solid black line with a slope of 1.78 $\pm$ 0.07. The grey line presents the K--S law with a threshold of 100 M$_{\solar}$ pc$^{-2}$. The slope of the grey line is 2.27 $\pm$ 0.21. The K--S relation with low $\Sigma _{\mathrm{H_{2}}}$ ($^{12}$CO, $X_{\mathrm{CO,v}}$) is overlaid with a thin solid black line.    
}
\label{FIG_KS13}
 \end{center}
\end{figure}

With the surface density derived in \S\ref{KS13_high},  a K--S plot in terms of $^{13}$CO is displayed in Figure \ref{FIG_KS13}. 
In this domain, $\Sigma _{\mathrm{H_{2}}}$ ranges from approximately 30 -- 400 M$_{\solar}$ pc$^{-2}$.  $\tau _{\mathrm{dep}}$ (or SFE) changes by more than one order magnitude over this range, from $\sim$ 1.5 $\times$ 10$^{9}$ yr at $\Sigma _{\mathrm{H_{2}}}$ $\approx$ 30 M$_{\solar}$ pc$^{-2}$ to $\sim$10$^{8}$ yr at $\Sigma _{\mathrm{H_{2}}}$ $\approx$ 400 M$_{\solar}$ pc$^{-2}$.
$N$ of the K--S plot is derived as 1.78 $\pm$ 0.07.

The best fit of the low-$\Sigma _{\mathrm{H_{2}}}$ regions based on $^{12}$CO ($X_{\mathrm{CO,v}}$) is over-plotted in Figure \ref{FIG_KS13}. It can be seen that the $\Sigma _{\mathrm{SFR}}$ of high-$\Sigma _{\mathrm{H_{2}}}$ regions lie beyond the best fit of low-$\Sigma _{\mathrm{H_{2}}}$ regions, implying a higher SFR at a given $\Sigma _{\mathrm{H_{2}}}$. 
\subsubsection{K--S Law at the High-$\Sigma _{\mathrm{H_{2}}}$ Based on $^{12}$CO} 
\label{KS12Var} 
The corresponding pixels in the $^{12}$CO map with significant $^{13}$CO emissions are used to create the K--S plot at high-$\Sigma _{\mathrm{H_{2}}}$ region for the comparison.

\begin{figure*}
\begin{center}
\FigureFile(80mm,80mm){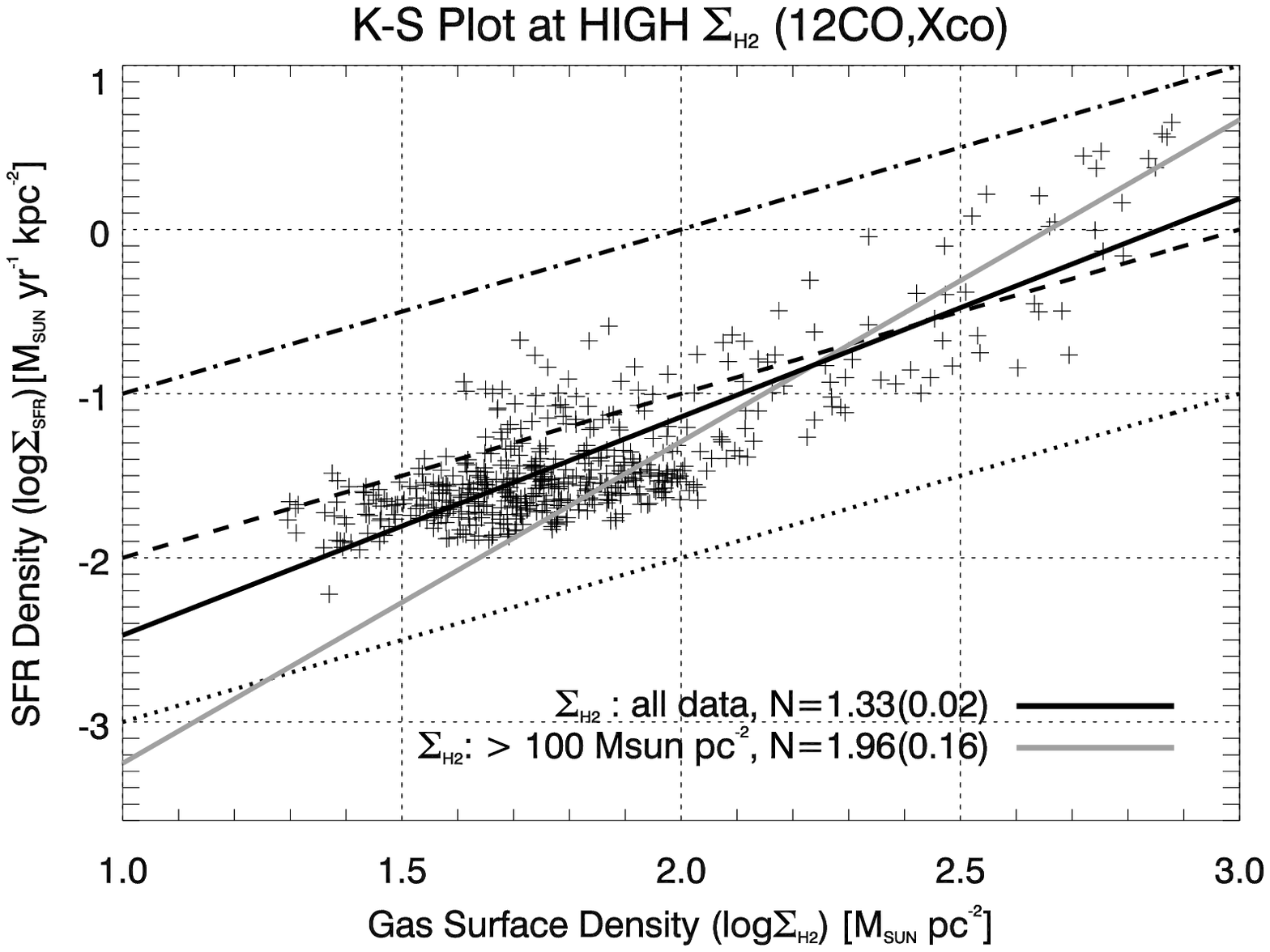}
\FigureFile(80mm,80mm){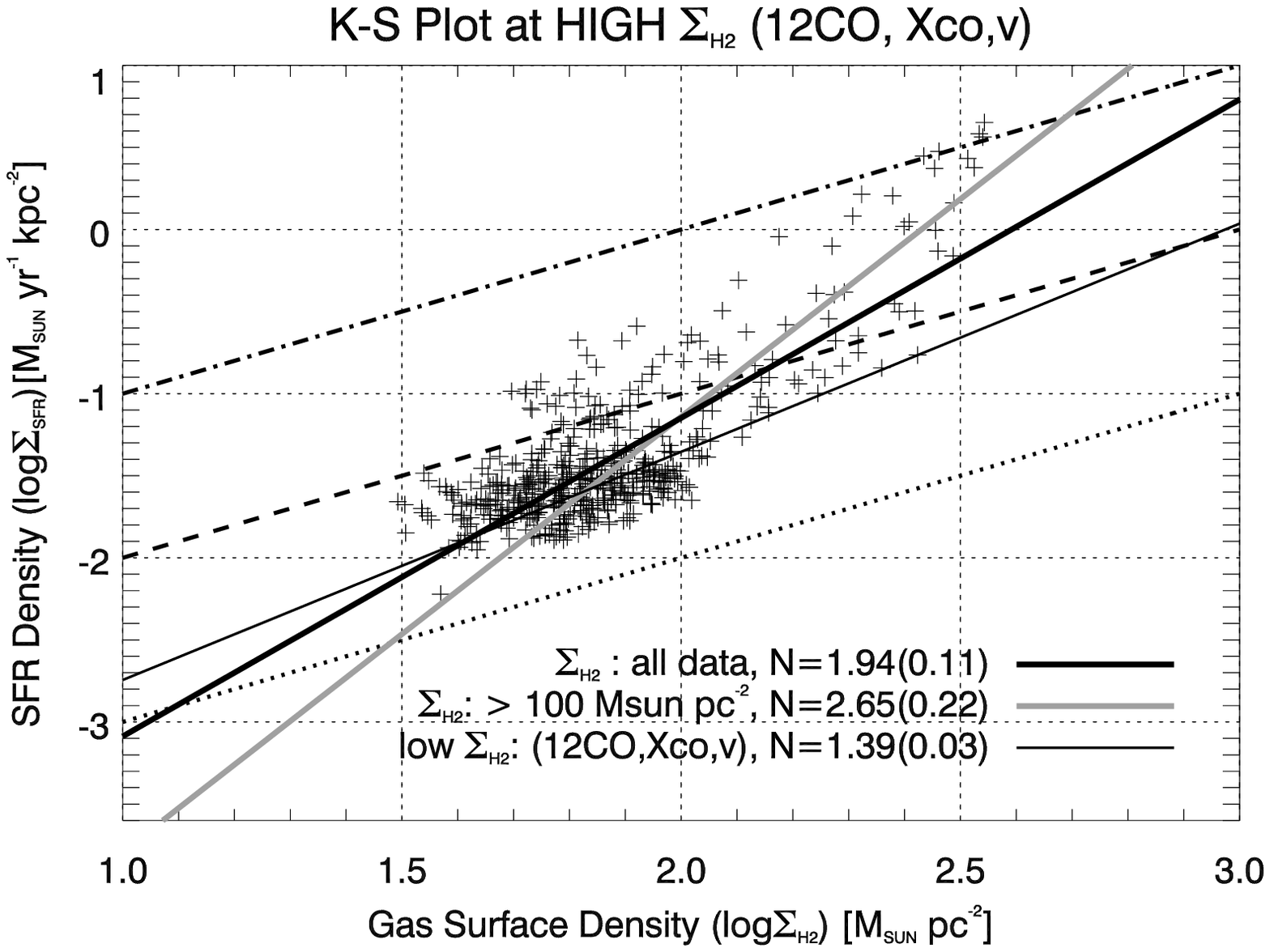}
\end{center}
\caption{K--S plots derived in high-$\Sigma _{\mathrm{H_{2}}}$ regions from $^{12}$CO. The colors and the style of the lines have the same meaning as in Figure \ref{FIG_KS13}. Left: K--S plot at high $\Sigma _{\mathrm{H_{2}}}$ based on $X_{\mathrm{CO}}$. Right: K--S plot at high $\Sigma _{\mathrm{H_{2}}}$ based on $X_{\mathrm{CO,v}}$.
}
\label{FIG_KS12}
\end{figure*}

First, the K--S plot is created with the Galactic $X_{\mathrm{CO}}$. The result is displayed in the left panel of Figure \ref{FIG_KS12}. 
$\Sigma _{\mathrm{H_{2}}}$ lies between 20 to 800 M$_{\solar}$ pc$^{-2}$. 
$N$ is 1.33 $\pm$ 0.02 with all data, slightly larger than the slope in the low-$\Sigma _{\mathrm{H_{2}}}$ region ($^{12}$CO, $X_{\mathrm{CO}}$).
The K--S plot with $X_{\mathrm{CO,v}}$ is shown in the right panel of Figure \ref{FIG_KS12}. The derived $N$ is approaching 2 (1.94 $\pm$ 0.11) with all data points.
$N$ derived in this section are larger than those in the environment of low-$\Sigma _{\mathrm{H_{2}}}$, regardless of whether  $X_{\mathrm{CO}}$ or  $X_{\mathrm{CO,v}}$ is used.
The variation of the power law index of the K--S law in this Section is summarized in Table \ref{TAB_KSslpoes}. 

\subsection{The Environmental Variation of the K--S Law}
\label{sec_environ_var}
There are two conclusions drawn from the above K-S plots. 
First of all, the adoption of $X_{\mathrm{CO,v}}$ steepens the K--S plots in both low- and high-$\Sigma _{\mathrm{H_{2}}}$ regimes. This is reasonable to expect from the difference in $X_{\mathrm{CO}}$ and $X_{\mathrm{CO,v}}$ (Figure \ref{FIG_XcoXcov}).
Secondly, there is a trend that $N$ would increase with increasing $\Sigma _{\mathrm{H_{2}}}$. 
The steeper slope in the higher-$\Sigma _{\mathrm{H_{2}}}$ area is confirmed by the independent measurement of $^{13}$CO. 

\begin{figure*}
\begin{center}
\FigureFile(80mm,80mm){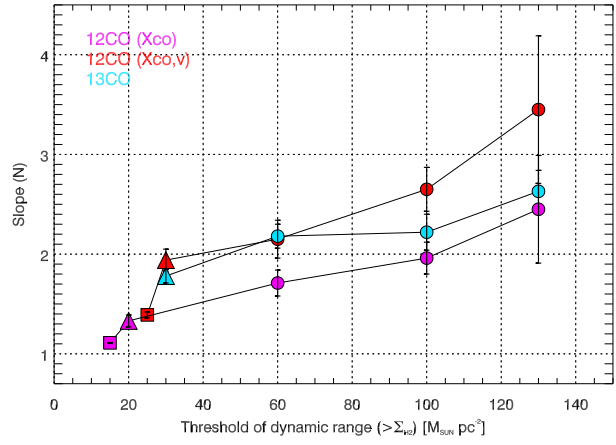}
\end{center}
\caption{Increasing slope of K--S law along $\Sigma _{\mathrm{H_{2}}}$. X-axis is the dynamic range of fitting and the y-axis is the K--S slopes ($N$). The data points with color of magenta, red, and cyan represent $\Sigma _{\mathrm{H_{2}}}$ from $^{12}$CO ($X_{\mathrm{CO}}$), $^{12}$CO ($X_{\mathrm{CO,v}}$), and $^{13}$CO, respectively. The results using all available data in the dataset of low and high $\Sigma _{\mathrm{H_{2}}}$ are presented in squares and triangles, respectively. The x-axis for the \emph{all data} points correspond to the detection limits. The circles denote the results with various thresholds. }
\label{FIG_N_bound}
\end{figure*}

A quantitative analysis of the dependence of $N$ on the  $\Sigma _{\mathrm{H_{2}}}$ considered is displayed in Figure \ref{FIG_N_bound}.
Figure \ref{FIG_N_bound} displays the derived $N$ with a different lower threshold of the used data. Only the data greater than the defined threshold of $\Sigma _{\mathrm{H_{2}}}$ are used. In other words, the dynamic range of fitting is narrowed down from \emph{all data} to \emph{$\mathit{\Sigma _{\mathrm{H_{2}}}}$ $>$ 130 M$_{\solar}$ pc$^{-2}$}, with intervals of $>$ 60, 100, and 130 M$_{\solar}$ pc$^{-2}$. 
The analysis was performed on three datasets, $^{12}$CO ($X_{\mathrm{CO}}$), $^{12}$CO ($X_{\mathrm{CO,v}}$), and  $^{13}$CO. 
Even though  $N$ with higher thresholds suffer from large uncertainties because of the scatter of measurements and limited data points, there is a clear trend that the slopes rise with an increasing threshold. 
In the low-$\Sigma _{\mathrm{H_{2}}}$ region (squares), $N$ is generally less than two. 
In the environment of $\Sigma _{\mathrm{H_{2}}}$ $>$ 100 M$_{\solar}$ pc$^{-2}$, $N$ increases to around two to three. Such a phenomenon is observed in both transitions.

Various studies have concluded that the SFR is well correlated with the amount of dense gas. 
The most straightforward evidence is the linear correlation between $L'_{\mathrm{HCN}}$ and $L_{\mathrm{IR}}$ (or $\Sigma _{\mathrm{dense\;gas}}$--$\Sigma _{\mathrm{SFR}}$). The correlation is held from Galactic dense cores to extragalactic unresolved sources \citep{Gao04,Wu05,Liu12}. Simulations of turbulent molecular clouds show that the SFR is controlled by the density probability distribution function (PDF) (e.g., \cite{Kra03,Wad07}). The galactic turbulence can be driven by gravitational instability, shocks or interaction of galaxies. \citet{Kra03} simulates the PDF of different gas surface densities across $\sim$ 20 -- 2000  M$_{\solar}$ pc$^{-2}$. The result shows that the fraction of dense gas (i.e., sufficiently dense for star formation) increases with gas surface density (see Figure 3 of \cite{Kra03}). Such a trend shown by the simulations has already been observed in the molecular clouds of our Milky Way (e.g., \cite{Kai09}).
Hence, the larger $N$ in the higher-$\mathit{\Sigma _{\mathrm{H_{2}}}}$ regime indicated by low density tracers (e.g., CO) may be a reflection of the increasing dense gas fraction.
\section{Comparisons with Other Galaxies}
\label{galaxies_literatures} 
The results from the study of IC 342 show that the K--S law may not be identical within a galaxy, 
namely, the slope steepens up toward the high-$\Sigma_{\mathrm{H_{2}}}$ region, reflecting a shorter $\tau_{\mathrm{dep}}$ (or higher SFE) and perhaps a different star formation mechanism.  
We examine whether the results of IC 342 are particular to this galaxy or a general situation by using the published data of some nearby galaxies.

There are two selection criteria for sample galaxies.
At first, the galaxy has to be studied in terms of the metallicity based on the framework of the P-method \citep{Pil00,Pi01a,Pi01b,Pil05}. The candidates reaching this criterion are shown in \citet{Pil04} and \citet{Mou10}. \citet{Pil04} and \citet{Mou10} derive the metallicity gradient with HII regions for their sample galaxies. \citet{Mou10} also derive the metallicity of the nuclear and circumnuclear regions for some of their samples.  
Secondly, the galaxy must be studied for $\Sigma_{\mathrm{H_{2}}}$ and $\Sigma_{\mathrm{SFR}}$.
 $\Sigma_{\mathrm{H_{2}}}$ and $\Sigma_{\mathrm{SFR}}$ were compiled from literature and divided into two groups.
The first group includes ten galaxies with published radial $\Sigma_{\mathrm{HI}}$, $\Sigma_{\mathrm{H_{2}}}$, and $\Sigma_{\mathrm{SFR}}$ from \citet{Ler08}.The physical resolution of the observations is $\sim$800 pc. 
$\Sigma_{\mathrm{H_{2}}}$ and $\Sigma_{\mathrm{SFR}}$ with $>$3$\sigma$ detection are used. Furthermore, for a fair comparison with IC 342, only the radii with $\Sigma_{\mathrm{H_{2}}}$$>$$\Sigma_{\mathrm{HI}}$ are used, i.e., the H$_{2}$-dominated regions. With the above two criteria, some data points at the outskirts of the galaxies are eliminated. The galaxies (NGC 628, NGC 3184, NGC 3198, NGC 3351, NGC 3521, NGC 4736, NGC 5055, NGC 5194, NGC 6946, and NGC 7331) in the first group will be used as the samples of {\it galactic disks} with lower $\Sigma_{\mathrm{H_{2}}}$.
The second group comprises nine \emph{starburst galaxies} from \citet{Ken98}, representing the high-$\Sigma_{\mathrm{H_{2}}}$ domain.  Most of starburst samples have CO and SFR observations with sub-kpc resolution towards their circumnuclear disks.
In such active star forming regions, it is reasonable to assume $\Sigma_{\mathrm{H_{2}}}$ $\gg$ $\Sigma_{\mathrm{HI}}$, as observed in the center of IC 342. The metallicity of the starburst samples are from either the 'circumnuclear' metallicity in \citet{Mou10}, or the extrapolation of the gradient toward the center from \citet{Pil04}.  
The second group contains NGC 253, NGC 1097, NGC 2093, NGC 3034, NGC 3351, NGC 4736, NGC 5194, NGC 5236, and NGC 6946.

\begin{figure*}
 \begin{center}
  \includegraphics[width=0.8\textwidth]{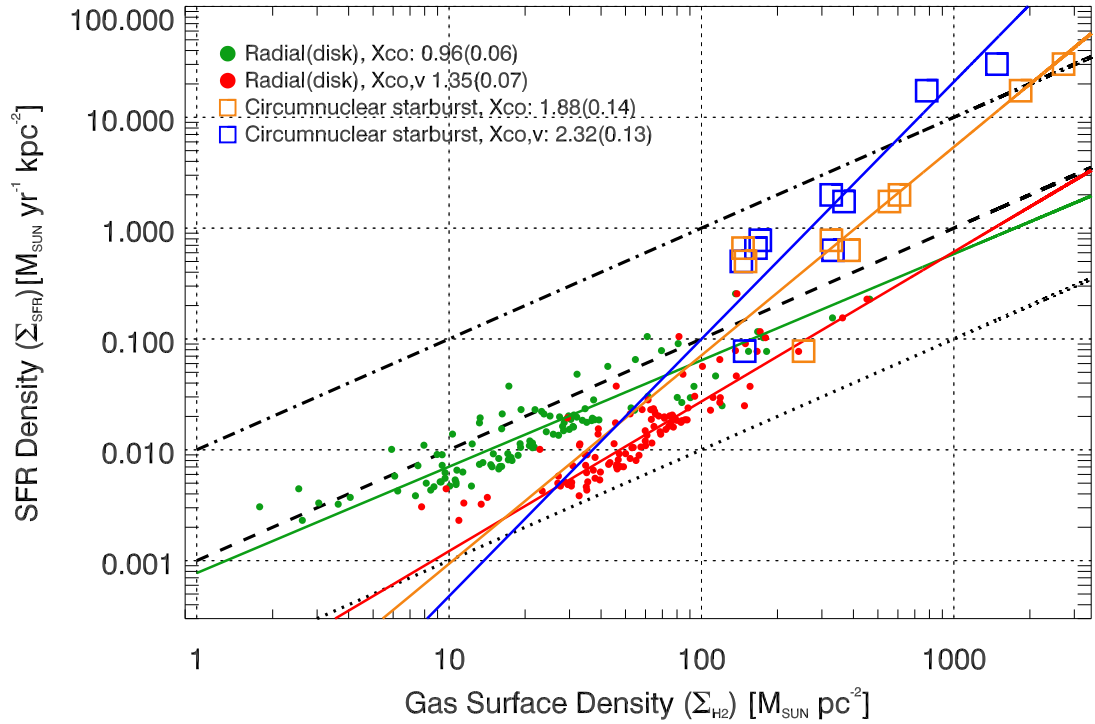}
  \caption{
K--S plot of nearby galaxies (exclude IC 342). The solid lines indicate the fitting results of the data points with the same color. Green circles: data from galactic disks with $\Sigma_{\mathrm{H_{2}}}$ from $X_{\mathrm{CO}}$ \citep{Ler08}. Red circles: data from galactic disks with $\Sigma_{\mathrm{H_{2}}}$ from $X_{\mathrm{CO,v}}$. Orange squares: data from starburst (SB) centers \citep{Ken98} with $\Sigma_{\mathrm{H_{2}}}$ from $X_{\mathrm{CO}}$. Blue squares: data from starburst (SB) centers with $\Sigma_{\mathrm{H_{2}}}$ from $X_{\mathrm{CO,v}}$. The $X_{\mathrm{CO}}$ $=$ 2.3 $\times$ 10$^{20}$ cm$^{-2}$ (K km s$^{-1}$)$^{-1}$ is used for the $\Sigma_{\mathrm{H_{2}}}$ based on Galactic $X_{\mathrm{CO}}$.
The derived slopes of the K--S law are shown in the legend. The error of each slope is shown in parentheses.
}
\label{FIG_ManyGalaxies}
 \end{center}
\end{figure*}

The K--S laws are examined on the basis of the Galactic Standard $X_{\mathrm{CO}}$ and $X_{\mathrm{CO,v}}$ calculated from the $^{12}$CO intensity and the metallicity.  
The results are shown in Figure \ref{FIG_ManyGalaxies}. 
$N$ for \emph{disk}, low-$\Sigma_{\mathrm{H_{2}}}$ samples with $X_{\mathrm{CO}}$ (green circles) is  0.96 $\pm$ 0.06 and  1.35 $\pm$ 0.07 for the same samples but using $X_{\mathrm{CO,v}}$ (red circles). 
Both values are consistent with the slopes of  IC 342 in the low-$\Sigma_{\mathrm{H_{2}}}$ regime (1.11 $\pm$ 0.02 and 1.39 $\pm$ 0.03, respectively).

For the circumnuclear starbursts, $\Sigma_{\mathrm{H_{2}}}$ is greater than 100 M$_{\solar}$ pc$^{-2}$, whether $X_{\mathrm{CO}}$ or $X_{\mathrm{CO,v}}$ is used.
$N$ is 1.88 $\pm$ 0.14 and 2.32 $\pm$ 0.13  by using $X_{\mathrm{CO}}$ and $X_{\mathrm{CO,v}}$, respectively. Both numbers are in good agreement with the value of $N$ derived from IC 342 with a threshold of 100 M$_{\solar}$ pc$^{-2}$, i.e., $N$ = 1.96 $\pm$ 0.16 and 2.65 $\pm$ 0.22 for $\Sigma_{\mathrm{H_{2}}}$ ($^{12}$CO, $X_{\mathrm{CO}}$) and  $\Sigma_{\mathrm{H_{2}}}$ ($^{12}$CO, $X_{\mathrm{CO,v}}$), respectively. 
The steeper $\Sigma_{\mathrm{H_{2}}}$--$\Sigma_{\mathrm{SFR}}$ relation toward high-$\Sigma_{\mathrm{H_{2}}}$ regime is not only found in local circumnuclear regions as this work shows, but also in more extreme objects (e.g., ULIRGs; \cite{Liu12}) and high redshift galaxies (e.g., \cite{Dec12}).
By comparing disk-averaged K-S law between ULIRGs and normal disk galaxies, \citet{Liu12} also suggest that K-S law is not uniform among different star forming populations. Because of their high SFE, $N$ is larger in ULIRGs than in normal disk galaxies.
Thus, we then have a certain level of confidence that the results of IC 342 are not particular but universal for at least nearby spiral galaxies.  
The slopes in this Section are summarized in Table \ref{TAB_KSslpoes} as well.
\section{Star Formation Mechanisms}
\label{SF_IC342}
The slope of the K--S plot depends on the mechanisms of star formation as mentioned above.
We discuss two possible mechanisms of star formation in IC 342,  gravitational instability and cloud-cloud collisions, as well as the theoretical aspects of changing the power law index of K--S law as a result of the properties of GMCs .    
\subsection{Gravitational Instability}
\label{Sec_grav}
For star formation caused by  gravitational instability, it is assumed that a constant fraction of molecular gas will be converted to stars per free fall time. Because $t_{\mathrm{ff}}=\sqrt{3\pi /32G\rho _{\mathrm{gas}}}$ and $\rho _{\mathrm{SFR}}\propto \rho _{\mathrm{gas}}/t_{\mathrm{ff}}$, where $\rho _{\mathrm{gas}}$ and $\rho _{\mathrm{SFR}}$ are gas mass and SFR per unit volume, $\rho _{\mathrm{SFR}}$ is proportional to $\rho _{\mathrm{gas}}^{1.5}$, or $\Sigma  _{\mathrm{SFR}}\propto \Sigma  _{\mathrm{gas}}^{1.5}$, assuming that the scale height of the gas disk is constant in galaxies. This naturally explains the well-known "disk-averaged" K--S law with a power law index of 1.4 in \citet{Ken98}. Our result of the K--S plot of 1.39 $\pm$ 0.03 in the low-$\Sigma _{\mathrm{H_{2}}}$ region (\S\ref{KSlowSig}) with $X_{\mathrm{CO,v}}$ is consistent with the slope expected by the theory and simulations of the collapse of a molecular cloud through gravitational instability \citep{Elm94,Elm02,Li06}.    

We then investigate the relation between the gravitational instability of gas and star formation in IC 342 via the Toomre Q parameter \citep{Too64} and 24$\mu$m image.  
\citet{Too64} suggests that the gas can collapse when $\Sigma  _{\mathrm{gas}}$ exceeds the critical density ($\Sigma  _{\mathrm{crit}}$), leading to a Toomre-Q parameter less than unity because
\begin{equation} 
\mathrm{\: Q}=\frac{\Sigma \mathrm{_{crit}}}{\Sigma _{\mathrm{gas}}},
\label{ToomreQ}
\end{equation} 
where $\Sigma  _{\mathrm{gas}}$ is $\Sigma  _{\mathrm{H_{2}+HI}}$.  $\Sigma  _{\mathrm{crit}}$ is determined by the epicyclic frequency $\kappa$ and velocity dispersion $\sigma$ as
\begin{equation} 
\Sigma _{\mathrm{crit}}=\frac{\kappa \sigma }{\pi \mathrm{G}}.
\label{SigCrit}
\end{equation}

An HI map of IC 342 was observed by \citet{Cro00} using the VLA (see \S\ref{sec_data_HI}). 
The rotation curve was derived with the Brandt model in \citet{Cro00} with the HI data. The maximum velocity of $\sim$170 km s$^{-1}$ is located at the radius of $\sim$5 kpc. $\kappa$ at a radius of $r$ is calculated by $\kappa^{2} (r)=2(\frac{v^{2}}{r^{2}}+\frac{v}{r}\frac{\mathrm{d} v}{\mathrm{d} r})$. The map of HI velocity dispersion \citep{Cro00} is used to estimate $\sigma$ in Equation (\ref{SigCrit}). $\sigma$ has a range approximately 7 -- 25 km s$^{-1}$.
To fairly derive the total gas mass at each pixel, the $^{12}$CO ($X_{\mathrm{CO,v}}$) and 24$\mu$m images were convolved to the same resolution as the HI image (38$\arcsec$ $\times$ 37$\arcsec$).

\begin{figure}
\begin{center}
\FigureFile(80mm,80mm){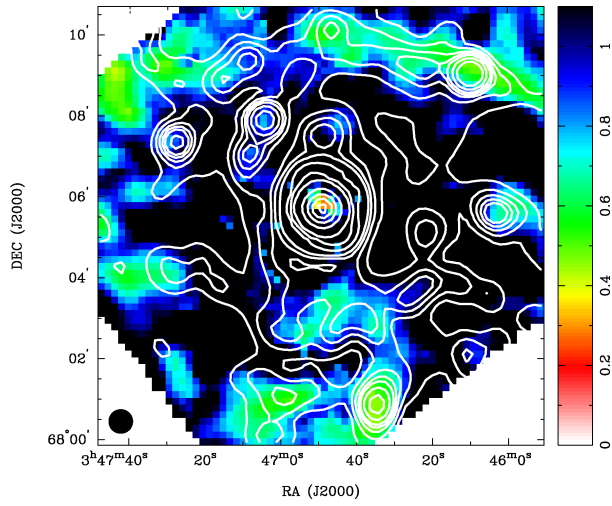}
\end{center}
\caption{The map of Tommre Q in a color scale overlaid with 24$\mu$m image in contours. The wedge indicates the corresponding values of Q parameter to the colors. The angular resolution in this map is 38$\arcsec$ to adjust with the HI map made with VLA. The synthesized beam is plotted in the lower left of the figure.
}
\label{FIG_ToomreQ}
\end{figure}

Firstly, the radial distribution of the Q parameter is displayed as a solid black line in Figure \ref{RadDist}. 
The radial profile shows that only at the galactic center Q  is $<$ 1. Outside the galactic center, the average Q parameter varies in a range of 1 -- 2. Because the gas distribution of IC 342 is not symmetric at each position angle of the galaxy, the radial distribution may smooth the real distribution of the Q parameter. For this reason, we then conduct the spatial distribution of the Q parameter.  
Figure \ref{FIG_ToomreQ} shows the spatial distribution of the Toomre Q parameter in a color scale overlaid on the 24$\mu$m image in contours.
The regions prone to collapse (Q $<$ 1) are shown in color. 
In the  innermost region of the galaxy, the bar end, and several knots in the spiral arms, Q $\approx$ 0.5-0.8.
The region where Q $<$ 1 is in agreement with the peaks of 24$\mu$m, indicating that the stars in the disk are likely formed through gravitational instability. Indeed, \citet{Yan07} suggest that $>$60\% of massive young stellar objects are inhabiting in the regions where the gas is unstable.

\subsection{Star Formation via Cloud-Cloud Collisions}
\label{Sec_ccc}
Star formation is known to potentially occur in regions with efficient cloud-cloud collisions \citep{Tan00,Kod06}. The collision time scale is $t_{\mathrm{coll}}\propto 1/nv\sigma \propto 1/\Sigma _{\mathrm{gas}}$, where $n$ is the number density of clouds, $v$ is the relative velocity of the clouds, and $\sigma$ is the average cloud cross section. $\Sigma  _{\mathrm{SFR}}$ per collision time is then proportional to $\Sigma  _{\mathrm{gas}}^{2}$. 

\citet{Ho82} suggest that there are $\sim$ 100 molecular cores in the central region of IC 342 and the cores are colliding with each other. 
The origin of the collisions can be attributed to bar-induced orbital crowding.   
The shocks produced after the cloud-cloud collisions have been found using shock tracer molecules, e.g., CH$_{3}$OH and HNCO \citep{Mei05}. 
These observations highlight the region where cloud-cloud collisions are promoted: along the leading side of the bar and the intersections of the bar and the circumnuclear ring (these two regions are located in the region defined as the center in Figure \ref{FIG_structure}, with a diameter of $\sim$ 1$\arcmin$; \cite{Mei05}). Indeed, both regions are associated with dense gas (e.g., HCN, \cite{Dow92}) and star forming tracers, such as H$\alpha$, and IR/radio star forming regions (e.g., \cite{Bec80,Tur83,Hir10}). 

Even though star formation by cloud-cloud collisions is suggested by previous observations with various lines, the possibility of gravitational instability cannot be ruled out because the Tommre Q parameter is well below unity in the central region of the galaxy (Figure \ref{FIG_ToomreQ}).The two mechanisms may work together \citep{Ran90}. That is, the gravitationally bound molecular clouds ($\sim$10$^{7}$ M$_{\solar}$) are formed through the gravitational instability; at the same time,  cloud-cloud collisions are taking place between the clouds to form giant molecular cloud associations (GMAs). This scenario is able to explain the high SFE of the spiral arm in M51 \citep{Ran90}. If the above scenario take places in the high-$\Sigma  _{\mathrm{H_{2}}}$ region of IC 342, $N$ = 2 -- 3 in this region represents a mix of the mechanisms.

\subsection{GMCs Properties and  Star Formation Law}
\label{Sec_TWOcomp}
In addition to the difference in the effective star formation mechanism, the K--S law may be altered because of the change in the intrinsic properties of GMCs.  

\citet{Kru09} have performed theoretical calculations of a two-component star formation law regardless of the galactic-scale process. The transition of $\Sigma  _{\mathrm{H_{2}}}$ is approximately 85 M$_{\solar}$ pc$^{-2}$.
In the regions where the $\Sigma  _{\mathrm{H_{2}}}$ is greater than $\sim$85 M$_{\solar}$ pc$^{-2}$, the galactic ISM pressure is sufficiently large to be comparable with the internal pressure of GMCs, and thus the density of GMCs is forced to increase with growing $\Sigma  _{\mathrm{H_{2}}}$ to balance the pressures. At the same time, the free-fall time of the GMCs is reduced \citep{Kru09}, resulting in faster star formation.  On the other hand, at low-$\Sigma  _{\mathrm{H_{2}}}$ regions, GMCs are independent of the environment so the $\Sigma  _{\mathrm{SFR}}$ is determined only by the GMCs themselves, implying a flattened K--S relation.


\section{Summary}
\label{sec_summary} 
We studied the Kennicutt-Schmidt (K--S) law in the nearby barred spiral galaxy IC 342 with published $^{12}$CO (1--0), $^{13}$CO (1--0), and infrared data. 
The main results are summarized as follows: 
\begin{enumerate}
\item After correcting for the oxygen abundance and $^{12}$CO intensity, the $^{12}$CO-to-H$_{2}$ conversion factor ($X_{\mathrm{CO,v}}$) was found to be  2--3 times lower than the Galactic standard $X_{\mathrm{CO}}$ in the  center of IC 342 and $\sim$2 times higher than the $X_{\mathrm{CO}}$ in the galactic disk (\S\ref{Sec_Xcolow}).
 
\item The power law index of the K--S plot  is approximately 1.4 in the low-$\Sigma_{\mathrm{H_{2}}}$ ($^{12}$CO,$X_{\mathrm{CO,v}}$) regions (often 25 -- 60 M$_{\solar}$ pc$^{-2}$), and increases to approximately 2 -- 3 at high-$\Sigma_{\mathrm{H_{2}}}$ ($^{12}$CO,$X_{\mathrm{CO,v}}$) regions  (around $>$  100 M$_{\solar}$ pc$^{-2}$) (\S\ref{sec_LargeSpatialKSXvar}, \S\ref{KS12Var}). The larger slope in the environment of high surface density is confirmed with independent measurement of $^{13}$CO (\S\ref{secKS13}). 

\item By setting different lower limits of the dynamic range for fitting, we confirmed that the slope of the K--S law is gradually steepened with increasing $\Sigma_{\mathrm{H_{2}}}$, presumably as a result of the increase in the dense gas fraction (\S\ref{sec_environ_var}).  

\item Using the published data of $\Sigma_{\mathrm{H_{2}}}$, $\Sigma_{\mathrm{SFR}}$ (measured at the sub-kpc scale), and the metallicity of nearby galactic disks and starburst centers, we confirm that the environmental variation in the K--S law in IC 342 may be a general case for spiral galaxies (\S\ref{galaxies_literatures}).

\item The varied slopes from $\sim$1.4 in the low-$\Sigma_{\mathrm{H_{2}}}$ domain to $\sim$ 2 -- 3 in the central region may indicate the change in the main mechanism of star formation among the sub-regions in IC 342, that is, the star formation is triggered by gravitational instability in the disk; at the galactic center, the combination of gravitational instability and cloud-cloud collisions is favored (\S\ref{Sec_grav} and \S\ref{Sec_ccc}). 

\item The derived variable K--S law in a single galaxy also matches the theoretical prediction that the GMC properties change at the $\Sigma_{\mathrm{H_{2}}}$ of approximately 85 M$_{\solar}$ pc$^{-2}$ and star formation efficiency increases with $\Sigma_{\mathrm{H_{2}}}$ (\S\ref{Sec_TWOcomp}). 

\end{enumerate}

\section*{Acknowledgements}
We would like to thank the anonymous referee for comments that helped to improve the manuscript.
This research has made use of the NASA/IPAC Extragalactic Database (NED) which is operated by the Jet Propulsion Laboratory, California Institute of Technology, under contract with the National Aeronautics and Space Administration.

\clearpage

\clearpage

\clearpage

\clearpage

\clearpage

\clearpage

\clearpage

\clearpage

\clearpage

\clearpage

\clearpage

\clearpage

\clearpage

\clearpage

\begin{center}
\begin{table*}
\begin{threeparttable}
\caption{Information on the original infrared data from archives.} 
  \normalsize{
\begin{tabular}{l*{6}{c}r}
\hline
\hline
Wavelength ($\mu$m)             & 24\tnote{a} & 70\tnote{b} &100\tnote{b}   &160\tnote{b}   & 250\tnote{c} & 350\tnote{c}   & 500\tnote{c}  \\
\hline
FWHM(arcsec)     & 5.7                      &  5.2                  &  7.7                     & 12                  &  18            & 25       & 37  \\
Pixel size (arcsec)         & 2.5                     & 1.4                      & 1.7                      & 2.85                &    6           &    10      & 14       \\
Image Unit                  & MJy/sr                 &  Jy/pixel               &  Jy/pixel               &  Jy/pixel         & MJy/sr  &  MJy/sr  &  MJy/sr  \\
\hline
\hline
\end{tabular}
}
\begin{tablenotes}
\footnotesize{
\item [a]Spitzer
\item [b]Herschel PACS
\item [c]Herschel SPIRE
}
\end{tablenotes}
 \label{IR_instrument}
\end{threeparttable}
\end{table*}
\end{center}

\clearpage

\begin{table*}
\centering
\caption{The slopes of K--S law derived in \S\ref{KS} and \ref{galaxies_literatures}}.
\begin{tabular}{l*{6}{c}r}
\hline
\hline
                                                     & $^{12}$CO ($X_{\mathrm{CO}}$) & $^{12}$CO ($X_{\mathrm{CO,v}}$) & $^{13}$CO\\ 
 \hline 
\multicolumn{4}{c}{{\bf IC 342}} \\
\hline
low $\Sigma _{\mathrm{H_{2}}}$ (all data)          &  1.11 $\pm$ 0.02  &  1.39 $\pm$ 0.03                        &   \dots  \\
high $\Sigma _{\mathrm{H_{2}}}$ (all data)   &   1.33 $\pm$ 0.02 &  1.94 $\pm$ 0.11 &   1.78 $\pm$ 0.07 \\
high $\Sigma _{\mathrm{H_{2}}}$ ($>$ 100 M$_{\solar}$ pc$^{-2}$)   &   1.96 $\pm$ 0.16 &  2.65 $\pm$ 0.22 &   2.27 $\pm$ 0.21 \\
\hline
\multicolumn{4}{c}{{\bf Spiral Galaxies (exclude IC 342)}} \\
\hline
low $\Sigma _{\mathrm{H_{2}}}$ (disk)  &         0.96 $\pm$ 0.06            & 1.35 $\pm$ 0.07   & \dots \\
high $\Sigma _{\mathrm{H_{2}}}$ (starburst)  &       1.88 $\pm$ 0.14                 & 2.32 $\pm$ 0.13   &  \dots\\
\hline
\hline
\end{tabular}
 \label{TAB_KSslpoes}
\end{table*}


\clearpage
\begin{thebibliography}{99}
\bibitem[Aalto et al.(2010)]{Aal10}Aalto, S., Beswick, R., J\"{u}tte, E. 2010, \aap, 522, 59 
\bibitem[Aniano et al.(2011)]{Ani11} Aniano G., Draine B. T., Gordon K. D., Sandstrom K., 2011, \pasp, 123, 1218
\bibitem[Arimoto et al.(1996)]{Ari96} Arimoto N., Sofue Y., \& Tsujimoto T., 1996, \pasj, 48, 275
\bibitem[Becklin et al.(1980)]{Bec80} Becklin, E. E. et al. 1980, \apj, 236, 441
\bibitem[Bergin et al.(1996)]{Ber96} Bergin, E. A., Snell, R. L., \& Goldsmith, P. F., 1996, \apj, 460, 343
\bibitem[Bernard et al.(2010)]{Ber10} Bernard, J.-Ph. et al. 2010, \aap, 518L, 88
\bibitem[Bigiel et al.(2008)]{Big08} Bigiel, F. et al. 2008, \aj, 136, 2846
\bibitem[Calzetti et al.(2005)]{Cal05} Calzetti, D. et al. 2005, \apj, 633, 871
\bibitem[Calzetti et al.(2007)]{Cal07} Calzetti, D. et al. 2007, \apj,666, 870
\bibitem[Calzetti et al.(2010)]{Cal10} Calzetti, D. et al. 2010, \apj, 714, 1256
\bibitem[Calzetti(2012)]{Ca12a}Calzetti D., 2012, preprint (arXiv:1208.2997)
\bibitem[Calzetti et al.(2012)]{Ca12b}Calzetti, D., Liu, G., \& Koda, J. 2012, \apj, 752, 98
\bibitem[Cedr\'{e}s et al.(2012)]{Ced12} Cedr\'{e}s, Cepa, J., Bongiovanni, \'{A}., Casta\~{n}eda, H.;,S\'{a}nchez-Portal, M., \& Tomita, A. 2012, \aap, 545, 43
\bibitem[Cox \& Mezger(1989)]{Cox89} Cox, P., \& Mezger, P. G. 1989, \aapr, 1, 49
\bibitem[Crosthwaite et al.(2000)]{Cro00} Crosthwaite, L. P., Turner, J. T., \& Ho P. T. P. 2000, \aj, 119, 1720
\bibitem[Crosthwaite et al.(2001)]{Cro01} Crosthwaite, L. P. et al. 2001, \aj, 122, 797
\bibitem[Decarli et al.(2012)]{Dec12}Decarli, R. et al. 2012, \apj, 752, 2
\bibitem[Dickman et al.(1986)]{Dic86} Dickman, R. L., Snell, R. L., \& Schloerb, F. P. 1986, \apj, 309, 326
\bibitem[Dobbs \& Pringle(2009)]{Dob09}Dobbs, C. L. \& Pringle, J. E. 2009, \mnras, 396, 1579

\bibitem[Downes et al.(1992)]{Dow92}Downes, D., Radford, S. J. E., Guilloteau, S., Guelin, M., Greve, A. \& Morris, D. 1992, \aap, 262, 424
\bibitem[Elmegreen(1994)]{Elm94} Elmegreen, B. G. 1994, \apj, 425, L73
\bibitem[Elmegreen(2002)]{Elm02}Elmegreen, B. G. 2002, \apj, 577.206
\bibitem[Espada et al.(2010)]{Esp10}Espada, D. et al. 2010, \apj, 720, 666
\bibitem[Feldmann et al.(2011)]{Fek11}Feldmann, R., Gnedin, N. Y., \& Kravtsov, A. V. 2011, ApJ, 732, 115
\bibitem[Galametz et al.(2012)]{Gal12} Galametz, M. et al. 2012, \mnras, 425, 763
\bibitem[Gao \& Solomon(2004)]{Gao04}Gao, Y., \& Solomon, P. M. 2004, \apjs, 152, 63
\bibitem[Genzel et al.(2012)]{Gen12} Genzel, R. et al. 2012, \apj, 746, 69
\bibitem[Gordon et al.(2008)]{Gor08} Gordon Karl D. et al. 2008, \apj, 682, 336
\bibitem[Goldsmith et al.(2001)]{Gol01} Goldsmith, Paul F. 2001, \apj, 557, 736
\bibitem[Goldsmith et al.(2008)]{Gol08} Goldsmith, P. F., Heyer, M., Narayanan, G., Snell, R., Li, D., \& Brunt, C. 2008, \apj, 680, 428
\bibitem[Henkel \& Mauersberger(1993)]{Hen93}Henkel, C.\& Mauersberger, R., 1993, \aap, 274,730
\bibitem[Henkel et al.(1998)]{Hen98}Henkel, C., Chin, Y.-N.; Mauersberger, R., Whiteoak, J. B. 1998, \aap, 329, 443

\bibitem[Hirota et al.(2010)]{Hir10} Hirota, A., Kuno, N., Sato, N., Nakanishi, H., Tosaki, T., \& Sorai, K. 2010, \pasj, 62, 1261

\bibitem[Ho et al.(1982)]{Ho82} Ho, P. T. P., Martin, R. N., \& Ruf, K. 1982, \aap, 113, 155
\bibitem[Israel(1997)]{Isr97}Israel, F. P. 1997, \aap, 328, 471
\bibitem[Keene et al.(1998)]{Kee98} Keene, J., Schilke, P. Kooi, J., Lis, D. C., Mehringer, D. M., \&  Phillips, T. G. 1998, \apj. 494, 107

\bibitem[Kennicutt(1998)]{Ken98} Kennicutt, R. C. Jr. 1998, \apj, 498, 541
\bibitem[Kennicutt e al.(2007)]{Ken07} Kennicutt, R. C. Jr. et al. 2007, \apj, 671, 333
\bibitem[Kennicutt e al.(2011)]{Ken11} Kennicutt, R. C. Jr. et al. 2011, \pasp, 123, 1347

\bibitem[Kainulainen e al.(2009)]{Kai09}Kainulainen, J., Beuther, H., Henning, T. \& Plume, R. 2009, \aap, 508L, 35
\bibitem[Koda \& Sofue(2006)]{Kod06} Koda, J. \& Sofue, Y. 2006, \pasj, 299, 312
\bibitem[Kravtsov(2003)]{Kra03}  Kravtsov, A. V. 2003, \apj, 590L, 1
\bibitem[Krumholz et al.(2008)]{Kru08} Krumholz, M. R., McKee, C. F., \& Tumlinson, J. 2008, \apj, 689, 865

\bibitem[Krumholz et al.(2009)]{Kru09} Krumholz, M. R., McKee, C. F. \& Tumlinson, J. 2009, \apj, 699, 850
\bibitem[Kuno et al.(2007)]{Kun07} Kuno, N. et al. 2007, \pasj, 59, 117
\bibitem[Leroy et al.(2008)]{Ler08}Leroy, A. K., Walter, F., Brinks, E., Bigiel, F., de Blok, W. J. G., Madore, B.; Thornley, M. D. 2008, \aj, 136, 2782
\bibitem[Li \& Draine(2001)]{Li01} Li, A. \& Draine, B. T., 2001, \apj, 554, 778
\bibitem[Li et al.(2006)]{Li06} Li, Y., Mac Low, M.-M \& Klessen, R. S. 2006, \apj, 639, 879
\bibitem[Liu et al.(2011)]{Liu11} Liu, G. et al. 2011, \apj, 735, 63
\bibitem[Liu \& Gao(2012)]{Liu12}Liu, L., \& Gao, Y. 2012, ScChG, 55, 347
\bibitem[Markwardt(2009)]{Mar09} Markwardt C. B., 2009, 2009, ASPC, 411, 251
\bibitem[Martel et al.(2012)]{Mar12} Martel, H., Urban, A., Evans, N. J., II 2012, \apj, 757, 59
\bibitem[McCall et al.(1985)]{McC85} McCall, M.L., Rybski, P.M. \& Shields G.A. 1985, \apjs, 57, 1
\bibitem[McQuinn et al.(2002)]{Mcq02} McQuinn, K. B. W.et al. 2002, \apj, 576, 274

\bibitem[Meier \& Turner(2001)]{Mei01} Meier, David S., Turner, Jean L. 2001, \apj, 551, 687
\bibitem[Meier \& Turner(2005)]{Mei05} Meier, David S. \& Turner, Jean L 2005, \apj, 618, 259
\bibitem[Milam et al.(2005)]{Mil05} Milam, S. N.; Savage, C.; Brewster, M. A.; Ziurys, L. M.; Wyckoff, S. 2005, ApJ, 634, 1126

\bibitem[Moustakas et al.(2010)]{Mou10} Moustakas, J., Kennicutt, R. C. Jr., Tremonti, C. A., et al. 2010, AJS, 190, 233
\bibitem[Nakai \& Kuno(1995)]{Nak95} Nakai, N., \& Kuno, N. 1995, \pasj, 47, 761
\bibitem[Narayanan et al.(2012)]{Nar12} Narayanan, Desika, Krumholz, Mark R., Ostriker, Eve C., \& Hernquist, Lars 2012, \mnras, 421, 3127
\bibitem[Narayanan et al.(2011)]{Nar11} Narayanan, D., Krumholz, M. R., Ostriker, E. C., \& Hernquist, L. 2011, \mnras, 418, 664
\bibitem[Onodera et al.(2010)]{Ono10} Onodera, S. et al. \apj, 2010, 722L, 127
\bibitem[Paglione et al.(2001)]{Pag01} Paglione, T. A. D., Wall, W. F., Young, J. S., et al. 2001, \apjs, 135, 183
\bibitem[Pineda et al.(2008)]{Pin08}Pineda, J. L. et al. 2008, \apj, 679, 481	
\bibitem[Pineda et al.(2010)]{Pin10}Pineda, J. L. et al. 2010, \apj, 721, 686
\bibitem[Pilyugin(2000)]{Pil00} Pilyugin, L. S. 2000, \aap, 362, 325
\bibitem[Pilyugin(2001a)]{Pi01a} Pilyugin, L. S. 2001a, \aap, 369, 594
\bibitem[Pilyugin(2001b)]{Pi01b} Pilyugin, L. S. 2001b, \aap, 373, 56
\bibitem[Pilyugin et al.(2004)]{Pil04} Pilyugin, L. S.; Vilchez, J. M.; Contini, T. 2004, \aap, 425, 849
\bibitem[Pilyugin et al.(2005)]{Pil05} Pilyugin, Leonid S.; Thuan, Trinh X. 2005, \apj, 631, 231
\bibitem[Pirogov et al.(2003)]{Pir03}Pirogov, L. et al. 2003, \aap, 405, 639
\bibitem[Popescu et al.(2002)]{Pop02} Popescu, C. C. et al. 2002, \apj, 567, 221

\bibitem[Rahman et al.(2011)]{Rah11} Rahman, N. et al. 2011, \apj, 730, 72
\bibitem[Rand \& Kulkarni(1990)]{Ran90} Rand, R. J. \& Kulkarni, S. R. 1990, \apj, 349, 43
\bibitem[Rieke et al.(2004)]{Rie04} Rieke, G. H., et al. 2004, \apjs, 154, 25
\bibitem[Rebolledo et al.(2012)]{Reb12}Rebolledo, D. et al. 2012, \apj, 757, 155 
\bibitem[Saha et al.(2002)]{Sah02} Saha, A., Claver, J., Hoessel, J. G.2002, \aj, 124, 839S
\bibitem[Schinnerer et al.(2003)]{Sch03}Schinnerer, E.,  B\"{o}ker, T., \&  Meier, D. S. 2003, \apj, 591L, 115
\bibitem[Schmidt(1959)]{Sch59}  Schmidt, M. 1959. \apj, 129,243
\bibitem[Schruba et al.(2010)]{Sch10} Schruba, A., Leroy, A. K., Walter, F., Sandstrom, K., \& Rosolowsky, E. 2010, \apj, 22, 1699
\bibitem[Strong et al.(1988)]{Str88} Strong, A. W., et al. 1988, \aap, 207, 1
\bibitem[Tasker \& Tan(2009)]{Tas09}Tasker, E. J. \& Tan, J. C.  2009, \apj, 700, 358
\bibitem[Tan(2000)]{Tan00} Tan, J. C. 2000, \apj, 536, 173
\bibitem[Toomre(1964)]{Too64}Toomre, A. 1964, \apj, 139, 1217
\bibitem[Turner \& Ho(1983)]{Tur83} Turner, J. L., \& Ho, P. T. P. 1983, \apj, 268, L79

\bibitem[Wada \& Norman(2007)]{Wad07} Wada, K., \& Norman, C. A. 2007, \apj, 660, 276
\bibitem[Watson et al.(1976)]{Wat76} Watson, W. D., Anicich, V. G., \& Huntress, W. T. 1976, \apj, 205, L165
\bibitem[Werk et al.(2011)]{Wer11} Werk J. K., Putman M. E., Meurer G. R., Santiago-Figueroa N., 2011, \apj, 735, 71
\bibitem[Wilson(1995)]{Wil95} Wilson, Christine D. 1995, \apj, 448L, 97
\bibitem[Wilson et al.(2009)]{Wil09} Wilson, T. L., Rohlfs, K., \& H\"{u}ttemeister, S. 2009, Tools of Radio Astronomy (Berlin: Springer-Verlag)

\bibitem[Wolfire et al.(2010)]{Wol10} Wolfire, M. G., Hollenbach, D., \& McKee, C. F. 2010, \apj, 716, 1191
\bibitem[Wu et al.(2005)]{Wu05}Wu, J. et al. 2005, \apj, 635L, 173
\bibitem[Yang et al.(2007)]{Yan07} Yang, C.-C., Gruendl, R. A., Chu, Y.-H., Mac Low, M.-M., \& Fukui, Y. 2007, \apj, 671, 374
\bibitem[Yao et al.(2006)]{Yao06} Yao, Lihong; Bell, T. A.; Viti, S.; Yates, J. A.; Seaquist, E. R. 2006, \apj, 636, 881

\end{thebibliography}
\end{document}